\documentclass[11pt]{article}
\usepackage{amsfonts}
\usepackage{amsmath, amsthm}
\usepackage{epsfig}
\usepackage{algorithm}
\usepackage{algorithmic}
\usepackage{epic}
\usepackage[a4paper]{geometry}
\usepackage{setspace}
\usepackage{enumitem}
\allowdisplaybreaks[4]

\usepackage{amsmath,verbatim,color,amssymb,epsfig}
\usepackage{bm}
\usepackage{booktabs}
\usepackage{ragged2e}
\usepackage{amsfonts}
\usepackage{epsfig}
\usepackage{changepage}
\usepackage{multirow}
\usepackage{makecell}
\usepackage{graphicx}
\usepackage{array}
\usepackage[table]{xcolor}
\usepackage[figuresleft]{rotating}
\usepackage{dcolumn}
\definecolor{Gray}{gray}{.80}
\usepackage{lscape}
\usepackage[round,semicolon,authoryear]{natbib}
\usepackage[normalem]{ulem}
\usepackage[colorlinks=true,urlcolor=blue,citecolor=purple,linkcolor=blue,bookmarks=true]{hyperref}
\usepackage{caption}
\usepackage{pdfsync}
\usepackage{sansmathaccent}
\usepackage{float}
\usepackage{diagbox}
\usepackage{subfigure}
\usepackage{supertabular}
\usepackage{threeparttable}
\begin{document}

\def\eqx"#1"{{\label{#1}}}
\def\eqn"#1"{{\ref{#1}}}

\makeatletter 
\@addtoreset{equation}{section}
\makeatother 

\def\yuancomment#1{\vskip 2mm\boxit{\vskip 2mm{\color{red}\bf#1} {\color{blue}\bf --Yuan\vskip 2mm}}\vskip 2mm}
\def\lincomment#1{\vskip 2mm\boxit{\vskip 2mm{\color{red}\bf#1} {\color{blue}\bf --Lin\vskip 2mm}}\vskip 2mm}
\def\squarebox#1{\hbox to #1{\hfill\vbox to #1{\vfill}}}
\def\boxit#1{\vbox{\hrule\hbox{\vrule\kern6pt
 \vbox{\kern6pt#1\kern6pt}\kern6pt\vrule}\hrule}}

\def\theequation{\thesection.\arabic{equation}}

\newcommand{\ds}{\displaystyle}

\newcommand{\bJ}{\mbox{\bf J}}
\newcommand{\bF}{\mbox{\bf F}}
\newcommand{\bM}{\mbox{\bf M}}
\newcommand{\bR}{\mbox{\bf R}}
\newcommand{\bZ}{\mbox{\bf Z}}
\newcommand{\bX}{\mbox{\bf X}}
\newcommand{\bx}{\mbox{\bf x}}
\newcommand{\bww}{\mbox{\bf w}}
\newcommand{\bQ}{\mbox{\bf Q}}
\newcommand{\bH}{\mbox{\bf H}}
\newcommand{\bh}{\mbox{\bf h}}
\newcommand{\bz}{\mbox{\bf z}}
\newcommand{\br}{\mbox{\bf r}}
\newcommand{\ba}{\mbox{\bf a}}
\newcommand{\be}{\mbox{\bf e}}
\newcommand{\bG}{\mbox{\bf G}}
\newcommand{\bB}{\mbox{\bf B}}
\newcommand{\bb}{\mbox{\bf b}}
\newcommand{\bA}{\mbox{\bf A}}
\newcommand{\bC}{\mbox{\bf C}}
\newcommand{\bI}{\mbox{\bf I}}
\newcommand{\bD}{\mbox{\bf D}}
\newcommand{\bU}{\mbox{\bf U}}
\newcommand{\bc}{\mbox{\bf c}}
\newcommand{\bd}{\mbox{\bf d}}
\newcommand{\bs}{\mbox{\bf s}}
\newcommand{\bS}{\mbox{\bf S}}
\newcommand{\bV}{\mbox{\bf V}}
\newcommand{\bv}{\mbox{\bf v}}
\newcommand{\bW}{\mbox{\bf W}}
\newcommand{\bY}{\mathbf{ Y}}
\newcommand{\bw}{\mbox{\bf w}}
\newcommand{\bg}{\mbox{\bf g}}
\newcommand{\bu}{\mbox{\bf u}}
\newcommand{\mI}{\mbox{I}}

\def\bb{{\bf b}}

\newcommand{\bcU}{\boldsymbol{\cal U}}
\newcommand{\bbeta}{\boldsymbol{\beta}}
\newcommand{\bdelta}{\boldsymbol{\delta}}
\newcommand{\bDelta}{\boldsymbol{\Delta}}
\newcommand{\boldeta}{\boldsymbol{\eta}}
\newcommand{\bxi}{\boldsymbol{\xi}}
\newcommand{\bGamma}{\boldsymbol{\Gamma}}
\newcommand{\bSigma}{\boldsymbol{\Sigma}}
\newcommand{\balpha}{\boldsymbol{\alpha}}
\newcommand{\bOmega}{\boldsymbol{\Omega}}
\newcommand{\btheta}{\boldsymbol{\theta}}
\newcommand{\bepsilon}{\boldsymbol{\epsilon}}
\newcommand{\bmu}{\boldsymbol{\mu}}
\newcommand{\bnu}{\boldsymbol{\nu}}
\newcommand{\bgamma}{\boldsymbol{\gamma}}
\newcommand{\btau}{\boldsymbol{\tau}}
\newcommand{\bTheta}{\boldsymbol{\Theta}}

\newtheorem{thm}{Theorem}
\newtheorem{lem}{Lemma}[section]
\newtheorem{rem}{Remark}[section]
\newtheorem{cor}{Corollary}[section]
\newcolumntype{L}[1]{>{\raggedright\let\newline\\\arraybackslash\hspace{0pt}}m{#1}}
\newcolumntype{C}[1]{>{\centering\let\newline\\\arraybackslash\hspace{0pt}}m{#1}}
\newcolumntype{R}[1]{>{\raggedleft\let\newline\\\arraybackslash\hspace{0pt}}m{#1}}

\newcommand{\tabincell}[2]{\begin{tabular}{@{}#1@{}}#2\end{tabular}}

\baselineskip=24pt
\begin{center}
{\Large \bf DEMO: Dose Exploration, Monitoring, and Optimization Using
a Biological Mediator for Clinical Outcomes}
\end{center}
\begin{center}
{\bf Cheng-Han Yang$^1$, Peter F. Thall$^2$, Ruitao Lin$^{2*}$}
\end{center}

\begin{center}

$^1$Department of Biostatistics, The University of Texas Health Science Center at Houston,\\
 Houston, TX 77030, USA\\

$^2$Department of Biostatistics, The University of Texas MD Anderson Cancer Center,\\
Houston, Texas 77030, U.S.A.\\

{$^*$Corresponding author: rlin@mdanderson.org}
\vspace{2mm}

\end{center}


\noindent{Abstract.}
Phase 1-2 designs  provide a methodological  advance over phase 1 designs for dose finding
by using both clinical response and toxicity.  A phase 1-2 trial still may fail to select a truly optimal dose.
 because early  response is not a perfect surrogate for long term therapeutic success.
To address this problem,
a generalized phase 1-2 design  first uses a phase 1-2 design's components to identify a set of  candidate doses,  adaptively randomizes
 patients among the candidates, and after longer follow up
 selects  a dose to maximize long-term  success rate.
In this paper, we  extend this paradigm by proposing a design
that exploits an early  treatment-related, real-valued biological outcome, such as  pharmacodynamic activity or an immunological effect,
that may act as a mediator between  dose and  clinical outcomes, including tumor response, toxicity, and survival time.
We assume  multivariate dose-outcome models that include   effects
 appearing in  causal pathways from dose to the clinical outcomes.
Bayesian model selection  is used  to identify and eliminate biologically inactive doses.
At the end of the trial, a
therapeutically optimal  dose is chosen from the set of doses that are acceptably safe, clinically effective,  and biologically active to maximize restricted mean survival time.
Results of a  simulation study show that the proposed design may provide substantial improvements
 over  designs that ignore the biological variable.

\vspace{.5cm}


\section{Introduction}

The current paradigm for early-phase oncology trials is changing. Conventionally,
a phase 1 oncology trial chooses  a maximum tolerated dose (MTD) based   on a binary  indicator, $Y_T$,
of dose-limiting toxicity (DLT).
Phase 1 designs originally were developed for cytotoxic  agents,  for which
 a higher dose is more likely to provide a  tumor response, indicated by $Y_R$, because  it kills more cancer cells,
   but  also is more likely to cause  toxicity because it kills more normal cells. Some commonly used phase 1 designs include a variety of
 3+3 algorithms,   versions of the continual  reassessment method (CRM) \citep{Cheung:2011book}, a Bayesian logistic regression model  (BLRM) based method \citep{neuenschwander2015bayesian},  and   Bayesian optimal interval  (BOIN) designs \citep{liu2015bayesian,yuan2022model}.
A well known problem with   Phase 1 designs is that they tend to select doses that are higher than necessary
  for   targeted agents or immunotherapies  having biological mechanisms  of action for which  increasing  administered dose beyond a certain level, say $\tau_B$, does not  increase efficacy  \citep{postel2016challenges, wages2018design, brock2021more}.   This  may occur if  the metabolized  agent reaches a saturation level in the patient,
  so that increasing the administered dose beyond $\tau_B$ does not increase the clinical response rate, but  may cause a higher risk of toxicity.   Because phase 1 trials are small with DLT defined based on short term follow up,
  such undesirable effects may not be seen until much later in the treatment evaluation process,  possibly during a phase 3 trial or in a
 large patient population  following regulatory approval \citep{Shah2021},
  which forces practicing physicians to make {\it ad hoc} dose adjustments.

By using both response and toxicity for screening and identifying an optimal dose, phase 1- 2 designs provide a substantial improvement over phase 1 designs.
Early phase 1-2 designs were proposed by  \cite{gooley},    \cite{thall1998strategy},  \cite{braun2002bivariate}, and  \cite{thall2004dose},  who gave Bayesian rules for sequential dose selection and monitoring safety and futility.
Utility-based phase 1-2 designs have been proposed by \cite{thall2012adaptive},   \cite{lin2017stein},
    \cite{lin2020boin12}, and many others. Reviews are given by \cite{yuan2016bayesian} and  \cite{yan2018}.

 Phase 1-2 designs
 still may  fail to identify  truly optimal doses because
short-term  response, $Y_R$, and toxicity, $Y_T,$ are not  perfect surrogates for a later clinical outcome that
 characterizes long-term  treatment success.   In oncology, long-term success may be defined in terms of
a time-to-event variable,  $Y_S$, such as
  remission duration (RD) among early responders
 \citep{hamada2018surrogate, ritchie2018defining},
progression-free survival (PFS) time,  or overall survival (OS) time, evaluated over longer follow up
than what is used to evaluate $Y_R$ and $Y_T$.
 For example, an immunotherapy may extend survival by controlling cancer progression without causing significant early tumor shrinkage, and consequently  its efficacy can only be established in terms of long-term  PFS or OS time.
To address this problem,
 \cite{thallGen12} proposed  the family of {\it Generalized Phase 1-2}  (Gen 1-2)  designs, which  use
 $Y_R$ and $Y_T$ for screening and long term therapeutic success defined in terms of $Y_S$ evaluated over longer follow up to optimize dose or schedule.

In this  paper,
we propose an extension  of the Gen 1-2 paradigm that incorporates
 a real-valued biological outcome,  $Y_{B}$, measured shortly after dose administration,
that may be related to dose effects on the clinical outcomes $Y_T$, $Y_R$, and $Y_S$.
We assume that  $Y_T$   and  $Y_B$  are
evaluated over a short time period, corresponding to one or two courses of therapy. This is  followed by
evaluation  of $Y_R$, and later $Y_S$,  subject to administrative censoring.
We will exploit the role of $Y_{B}$  as a mediator between   dose $d$ and  $(Y_T, Y_R,Y_S)$.  This is
 illustrated by the directed acyclic graph (DAG)   in Figure \ref{fig:DAG}, which  shows that a dose
 $d$ may  influence each  of the outcomes $Y_T$,  $Y_R$, and $Y_S$  directly, and also through  indirect effects mediated by $Y_{B}.$   The DAG  also allows  interactions between $Y_T$ and $Y_R$.
The rationale for incorporating $Y_B$ into the dose-outcome model and design
is that, if $Y_B$  mediates  dose effects on the clinical outcomes,
the additional information provided by $Y_B$  may increase the probability of identifying a  truly optimal dose.
 To simplify the exposition,
in the sequel  we will refer to $Y_S$ as survival time.

The proposed design,  which we call DEMO, does  \underline{D}ose \underline{E}xploration, \underline{M}onitoring, and \underline{O}ptimization.
A DEMO design has three stages,  (1) sequential dose exploration,  including elimination of unsafe or biologically inactive doses, (2)  randomized dose screening, and (3) randomized dose optimization, with
successively more intensive monitoring as the trial progresses.
Because DEMO incorporates  $Y_B$,  it  is similar in spirit
to the dose-finding designs of  \cite{ursino2017dose} and \cite{gerard2022bayesian}, who  explored  methods  utilizing PK information, \cite{takeda2018bayesian}, who developed a PK exposure-toxicity model based on the area under the concentration (AUC) curve, and \cite{su2022semi}, who  proposed a  semi-mechanistic design based on dynamic PK/PD modeling.

In addition to incorporating $Y_B$, another  important difference between  DEMO and  a Gen 1-2 design is the criterion used for optimal dose  selection.
In DEMO, an optimal
therapeutic dose (OTD) is defined as the dose that maximizes restricted mean survival time  (RMST),
among doses that are acceptably safe, have an acceptable clinical response rate,  and
are biologically active.  In contrast, a Gen 1-2 design relies on  mean survival time (MST) or
$\Pr(Y_S > t_S  \,|\, d)$ for given fixed $t_S$, such as six months, as a criterion for selecting an optimal dose.
An advantage of RMST over MST is that  RMST may be evaluated at any meaningful time point. Early-phase trials often have limited follow-up time
 that may not be long enough to allow MST to be estimated reliably. In such cases, RMST still can be computed. This flexibility facilitates analysis of $Y_S$ for a  variety of different timeframes. In practice, the estimated RMST  often is more reliable than the estimated mean or median survival time  \citep{royston2013restricted}.

\section{The DREAMM Studies}

To  motivate the DEMO design,  we  review a sequence of clinical trials conducted to evaluate the safety and effectiveness of belantamab mafodotin (GSK2857916) for  patients with relapsed or refractory multiple myeloma  \citep{trudel2018targeting, lonial2020belantamab}. Belantamab mafodotin  is an immunoconjugate that targets the B-cell maturation antigen (BCMA), a tumor necrosis superfamily of cell-surface receptors  required for  cell survival. In the DREAMM-1 trial (NCT02064387),  79 patients were assigned sequentially  to  doses in the set
 \{0.03, 0.06, 0.12, 0.24, 0.48, 0.96, 1.92, 2.5, 3.4,  4.5\} mg/kg, using the
  BLRM,
with  a target toxicity rate  of 25\%.  No MTD was reached, and 3.4 mg/kg was identified as the recommended phase 2 dose (RP2D).
  The bioactivity of  belantamab mafodotin may be
  quantified by the free soluble BCMA level measured after  infusion, defined as $Y_B$ = $-$log([post-infusion BCMA]/[baseline BCMA]) evaluated at day 7,
  with  larger   $Y_B$ corresponding to greater biological activity.  The overall response rate (ORR) and PFS time were evaluated to characterize clinical efficacy.



While    3.4 mg/kg achieved the highest ORR   and no DLTs were observed in DREAMM-1,
a substantial proportion of patients needed subsequent dose reductions  to manage  later adverse events (AEs).   This motivated   the investigators to conduct the  DREAMM-2 trial (NCT03525678), in order to generate additional safety and efficacy data at the RP2D and compare it to  a lower dose.   In DREAMM-2,
patients were randomized  fairly between 2.5 mg/kg ($n$ = 97) and 3.4 mg/kg ($n$ = 99).
The DREAMM-2 data showed no clinically meaningful efficacy differences, with  ORR  rates 31\% at 2.5 mg/kg and 34\% at 3.4 mg/kg, and  estimated  median PFS  11.0 months
for  2.5 mg/kg  and 13.7 months
for   3.4 mg/kg.
Since the 2.5 mg/kg cohort had a slightly smaller severe AE rate of 40\%  versus 47\% for 3.4 mg/kg, the RP2D  was chosen to be 2.5 mg/kg because it showed comparable anti-myeloma activity and a more favorable safety profile.  However, if PFS had been used as the primary efficacy endpoint
3.4 mg/kg would have been selected because it  had estimated median PFS  2.7 months longer than that of   2.5 mg/kg.
A  more recent trial was DREAMM-3,  a phase 3 randomized trial comparing belantamab mafodotin monotherapy given at  2.5 mg/kg to the combination of pomalidomide and dexamethasone. Using PFS time  as the primary endpoint, the outcome of the DREAMM-3 trial was negative.

Considered together, the DREAMM trials illustrate what can go wrong in settings where dose exploration and optimization are carried out in separate trials, and a key biological outcome that is related to clinical outcomes  is not incorporated into the study design. In general, organizing and conducting a sequence of  trials in an {\it ad hoc} manner may lead to increased overall trial duration,  higher costs, and suboptimal decisions.
For example, the choice of the two doses compared in  DREAMM-2  was not formally justified, and cannot serve as a model for designing future trials.   The DREAMM-2 study also showed a disagreement between estimates of the intermediate outcome OR and long-term PFS, which in general is not uncommon.

The   DEMO design addresses practical and technical challenges that arose in the DREAMM studies. DEMO is a seamless three-stage design that eliminates the  gap between separate trials and, ideally, might have been used
to construct a single trial to replace the two DREAMM trials.
Throughout its three stages, DEMO employs multiple endpoints, including short-term bioactivity and toxicity, an intermediate tumor response indicator,
and  long-term   survival.  Together, the three stages of DEMO provide a coherent basis for making better informed decisions based on the available data,  and increasing the probability  that the chosen dose will be safe and
maximize  long-term survival.  Because  DEMO  selects an optimal dose based on long-term survival  data, it aligns with the conventional goal of  randomized confirmatory trials.

\section{Dose Acceptability and Optimality Criteria}
\subsection{Notation}
While the final objective of DEMO is to select an optimal dose
based on RMST,   during the trial undesirable doses are dropped
using  acceptability criteria defined in  terms of $Y_T,$ $Y_B,$ $Y_R,$ and  $Y_S.$
Dropping doses determined to be unacceptable  based on interim data provides a basis for enriching the sample sizes
for acceptable doses that are potentially optimal, and also benefits future patients enrolled in the trial.
Denote the doses
 by $d_1 < d_2 < \cdots < d_J$ and, for brevity,  temporarily suppress notation for model parameters.  
   For simplicity, and to focus on the key elements of DEMO, we will assume that   $Y_R$  and $Y_T$ are binary indicators,
and denote $\pi_k(d_j)$
= $\Pr(Y_k=1  \,|\, d_j)$   for  $k=R,T.$   If desired, the DEMO design may be generalized 
to accommodate ordinal $Y_R$  and $Y_T,$ with appropriate model extensions.
  The density, survival probability function, and hazard function of $Y_S$
for a patient treated with dose $d_j$ are denoted, respectively,  by
$f_S(y  \,|\, d_j),$  ${\mathcal F}_S(y  \,|\, d_j)$, and $h_S(y  \,|\,  d_j)$ = $f_S(y  \,|\,  d_j)/{\mathcal F}_S(y  \,|\, d_j)$, for $y>0.$
Given  a fixed  follow up time $t_S$,  the RMST of $d_j$ is $\mu_{S}(d_j)$ = $\int_0^{t_S}{\mathcal F}_S(y  \,|\, d_j)dy.$
The DEMO design includes dose acceptability criteria defined in terms of  fixed  limits  elicited from  the clinical investigators, including
 $\overline{\pi}_T$ = the maximum acceptable $\pi_T(d_j)$,
$\underline{\pi}_R$ = the minimum acceptable $\pi_R(d_j)$,
 and $\underline{\mu}_S$ = the minimum acceptable $\mu_{S}(d_j)$.

\subsection{Biological Endpoints}
The main extensions of a Gen 1-2 design provided by DEMO arise from incorporating a biological endpoint, $Y_B$.
This  is used to screen out biologically inactive doses, and to inform the distributions of the clinical variables by regression on $Y_B.$
 Depending on the trial's setting, there are many  well-established  biological endpoints that are known to be related to clinical outcomes, and  thus they may be used as $Y_B$ in a DEMO design.
Examples of $Y_{B}$ include  pharmacokinetic (PK) or pharmacodynamic (PD) variables
such as the maximum serum concentration (Cmax), drug exposure characterized by pharmacologic area under
 the curve (AUC)  that characterizes systemic exposure of a metabolized agent, or minimum inhibitory concentration (MIC).
Other examples of $Y_B$ include the expression level of a  protein involved in a signaling pathway,
  a targeted biomarker of programmed cell death-ligand  such as PD-L1 expression level for a checkpoint inhibitor, a  B cell surface glycoprotein CD19,  the inflammatory cytokine IL-15,
 and cell level following infusion of engineered CAR-NK or T-cell therapies.
In the motivating DREAMM studies,  $Y_{B}$ would be
  BCMA level for belantamab mafodotin.   Well-established biomarkers  exist that are associated with response rate or long-term survival, such as the level of circulating tumor DNA (cfDNA) for lung cancer \citep{cisneros2022cell}.
For a cell therapy trial, \cite{walter2012multipeptide} reported that  T-cell responses were associated with better disease control and longer survival time for renal cell cancer patients. \cite{selitsky2019prognostic} found that B cell characteristics in skin cutaneous melanoma play an important  role in predicting $Y_R$ and $Y_S$  following immunotherapy.

Unlike design
parameters used for  clinical endpoints, such as a maximum probability of toxicity or minimum probability of response,  it typically is not feasible  to establish a target or threshold value for
$\mu_{B}(d_j)$ = $E(Y_B \,|\, d_j)$.   This is because,
  in practice, what may be desirable or undesirable values of $\mu_{B}(d_j)$ in terms of effects on clinical outcomes  typically have not been established,  especially in first-in-human trials.  Below, we will present a data-adaptive approach to identify and drop biologically inactive doses based on $Y_B$.  In the sequel, we will use the terms ``biological activity'', ``bioactivity'', and ``biological endpoint'' interchangeably to refer to $Y_B$.

\begin{figure}
  \begin{center}
\includegraphics[width=0.65\textwidth]{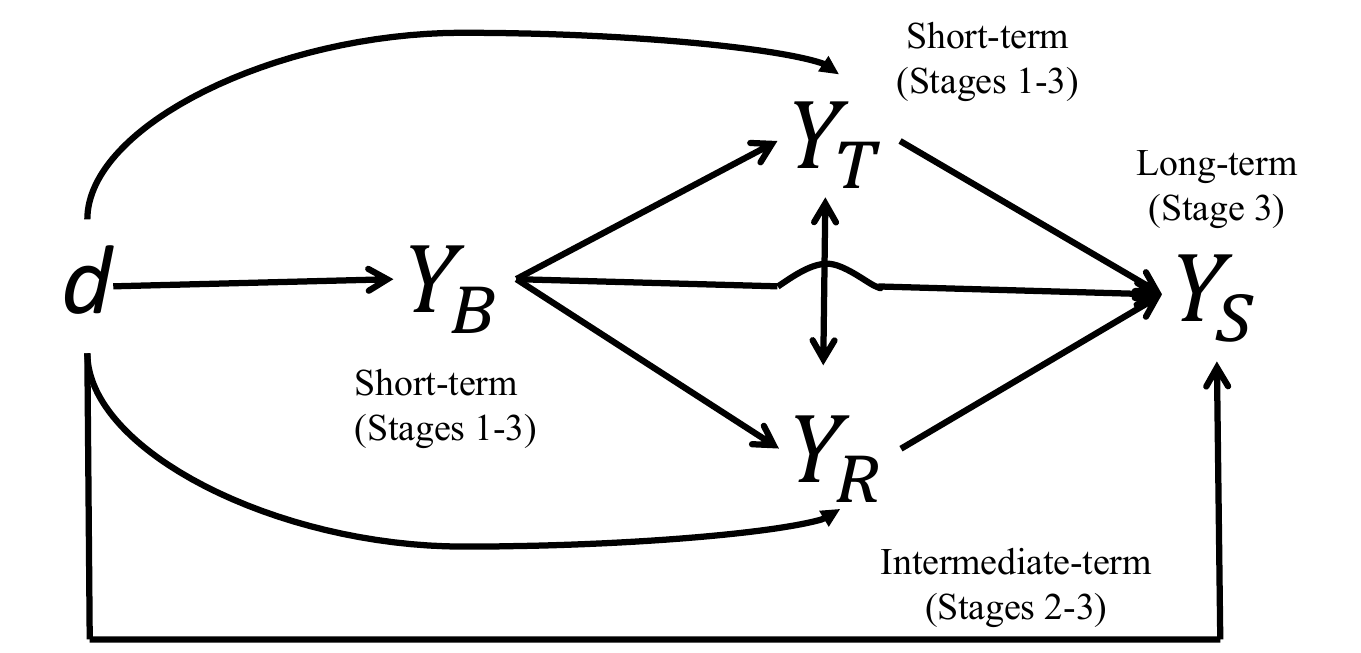}
\end{center}
 \caption{Causal diagram for dose $d$, biomarker $Y_{B}$, phase 1-2 early toxicity and response outcomes $Y_T$ and $Y_R$, and survival time $Y_S$.}
\label{fig:DAG}
\end{figure}

Rather than assuming that $Y_{B}$ is  a surrogate endpoint for either $Y_R$ or $Y_S,$
  we make the more realistic assumption
   that $Y_B$ may act as a mediator between
$d$ and the clinical  variables $(Y_R,Y_T,Y_S)$, as shown in Figure \ref{fig:DAG}.
While the distribution of $Y_{B}$ may be constructed to reflect a particular biological effect,
for tractability we make the practical assumption that
\begin{equation}\label{Bmodel}
Y_{B}  \,|\, d_j \ \sim\ \mathcal{N}(\mu_{B}(d_j),\ \sigma_{B}^2),
\end{equation}
where  $\mu_{B}(d_j)$ = $E(Y_B  \,|\, d_j)$ is assumed to increase with $d_j$.
This monotonicity assumption  reflects the underlying biology of the dose effect on $Y_B$ in most real-world applications. For example, a larger number of infused T-cells should lead to  a higher level of targeted biomarker engagement, quantified by $Y_B$.   In contrast,  a response indicator $Y_R$ or survival time $Y_S$ each may be impacted by many different factors apart from  dose.   If the underlying biology does not imply monotonicity of  $\mu_{B}(d_j)$, however, the model
for $Y_B$ and the methodology should be modified accordingly.
In the DEMO design,  at successive stages of the trial,  different
  models for  $\mu_B(d_j)$ are assumed for different purposes.  The first model, used in stage 1,  is constructed to provide a basis for screening out   biologically inactive doses.
  The second model is used to exploit causal relationships between $Y_B$ and the clinical outcomes (Figure \ref{fig:DAG}) when using the trial's final data  to identify the optimal dose.

We  identify a set of {\it  biologically inactive} doses by formulating  the first model for $\mu_B(d_j)$
as a  simple step function on the interval from 0 to $d_J$ that may have at most one
increasing step at one of the $d_j$'s.   We denote these step functions by
$\mu_B^*(d)$ for $d\in \{d_1,\ldots,d_J\}$,
to distinguish them from the general class of  mean functions $\mu_B(d).$
 Using the set of $\mu_B^*(d_j)$'s, we  focus on the problem of identifying the dose
in  $\{d_1, d_2, \cdots, d_J\}$ where the increasing step occurs, and we denote this dose by $\tau_B$.
We define the set of step functions to include the constant function having no increase,
with  $\mu_B^*(d_1) = \cdots = \mu_B^*(d_J)$, and in this case we define $\tau_B$ = $d_1$
and do not declare any doses to be biologically inactive.
The next step function has $\mu_B^*(d_1)  < \mu_B^*(d_2) = \cdots = \mu_B^*(d_J)$ with $\tau_B$ = $d_2,$
and so on, up to
the last step function where $\mu_B^*(d_{1})= \mu_B^*(d_{J-1})  < \mu_B^*(d_J)$ and $\tau_B$ = $d_J.$
For each of the
$J-1$ step functions with $\tau_B$ = $d_j$ for  $j=2,\cdots,J,$ we
define the lower doses $\{d_1,\cdots, \tau_B-1\}$  to be {\it biologically inactive}
and  the upper doses $\{\tau_B,\cdots, d_J\}$  to be  {\it biologically   active}.

Using this structure, we  formulate estimation of  $\tau_{B}$ as a  Bayesian model selection problem, with
$J$ possible models for $\mu_B^*(d),$ where each model corresponds to one of the step functions.
For each $j=1,\cdots,J,$  we define the $j^{th}$ model,
$\mathcal{M}_{j},$ to be  the step function for which $\tau_{B}=d_j$, and
denote its prior probability by   $\Pr(\mathcal{M}_{j})$.
To determine whether each dose is biologically inactive ($-$) or active ($+$) using this construction, the  distribution of $Y_{B}$ under
$\mathcal{M}_{j}$ is assumed to be  normal with  common variance  $\sigma^2_B$ and  mean that is one of two values, $\mu_{-}$ or  $\mu_{+} > \mu_{-}$.
Denoting the observed data by ${\mathcal D}$ and  priors for $\mu_{-}$, $\mu_+$, and $\sigma^2_B$ by $\pi(\mu_{-})$, $\pi(\mu_{+})$, and $\pi(\sigma_B^{2})$, respectively,
 the posterior of  $\mathcal{M}_{j}$ is
\begin{eqnarray*}
{\displaystyle \Pr(\mathcal{M}_{j} \,|\,\mathcal{D})} & \propto & \Pr(\mathcal{M}_{j})\int\pi(\sigma_{B}^{2})\left\{ \int\prod_{r=1}^{j-1}\phi_{r}(\mu_{-},\sigma_{B}^{2}\  \,|\,\ \mathcal{D})\pi(\mu_{-}){\rm d}\mu_{-}\right.\\
 &  & \left. \times\int\prod_{r=j}^{J}\phi_{r}(\mu_{+},\sigma_{B}^{2}\  \,|\,\ \mathcal{D})\pi(\mu_{+}){\rm d}\mu_{+}\right\} {\rm d}\sigma_{B}^{2}.
\end{eqnarray*}
 For each dose index $r=1,\cdots,J$ in this expression,  $\phi_{r}$ denotes
the  product likelihood of  normal pdfs of the observed $Y_B$ values of all patients  treated  at dose $d_{r}$. We assume priors
\[
\pi(\mu_{+})= \mathcal{N}(m_{\mu_{+}},v^2_\mu),\ \
\pi(\mu_{ -})=\mathcal{N}(m_{\mu_{-}},v^2_\mu),\  \ \pi(\sigma_{B}^{-2})= {\rm Gamma}(a_\sigma,b_\sigma),
\]
where $m_{\mu_{-}} < m_{\mu_{+}}.$  In the Supplementary Materials, we give elicited values of
the hyperparameters
$m_{\mu_-}$, $m_{\mu_+}$, $v^2_\mu$, $a_\sigma$ and $b_\sigma$, and  derive a closed form for  $\Pr(\mathcal{M}_{j} \,|\,\mathcal{D})$.  For simplicity, we assume a  discrete uniform prior
on the model probabilities,  $\Pr(\mathcal{M}_{j})=1/J$ for all $j$.

Using a model selection framework,
a cutoff $c_B$ is specified so that  a dose $d_j$ is considered  biologically inactive based on the data $\mathcal{D}$ if
$\Pr(\mathcal{M}_{r} \,|\, \mathcal{D} ) > c_B$ for a higher dose $d_r>d_j$.
  We  estimate the dose where the upward step occurs as
$\hat{\tau}_B$   =  $\hat{\tau}_B(\mathcal{D})$  =
the smallest dose $d_j$ such that $\Pr(\mathcal{M}_{j} \,|\, \mathcal{D})  > c_B$.
If   no dose satisfies this posterior criterion then $\hat{\tau}_B $ = $d_1$, and  no doses are declared biologically inactive.
If $\hat{\tau}_B > d_1$
then doses $\{d_1, \cdots, \hat{\tau}_B-1\}$ are declared biologically inactive and dropped.

\subsection{Dose Acceptability and Optimality}
The  OTD, which is the criterion used by DEMO for final  dose selection, is defined
 using all  four endpoints $Y_B, Y_T, Y_R,$ and  $Y_S$.  The definition is formulated to accommodate settings where the $RMST(d)$ curve increases to a plateau
 at a dose $\tau_B$,  and no longer increases for higher doses. In such as case, more than one dose in $\{d_1, \cdots, d_J\}$ may maximize  $RMST(d)$.

 \vskip.1in
\noindent {\sc Definition}  The  {\it Optimal Therapeutic Dose} (OTD) is the lowest dose  $d_j$  in $\{d_1,\ldots,d_J\}$ that is biologically active ($d_j\geq \tau_B$), safe ($\pi_T(d_j)< \overline{\pi}_T$), and clinically effective in  terms of both response probability ($\pi_R(d_j)>\underline{\pi}_R$) and survival time ($\mu_S(d_j)>\underline{\mu}_S$), that maximizes the RMST $\mu_S(d_j)$.
 \vskip.1in

The following four statistical criteria are used during the trial to determine which doses are acceptable.
Given observed data ${\mathcal D}$ and the decision cutoffs, $c_B,$ $c_T,$ $c_R,$  and $c_S$,  a dose $d_j$ is an {\it acceptable OTD candidate} if it satisfies all of the following posterior inequalities:
\begin{subequations}
\begin{align}
\hspace{-3cm} {\rm Biologically\ Active} & \hspace{2em}   d_j \geq   \hat{\tau}_B(\mathcal{D})  \\
 {\rm Safe}\ & \hspace{2em} \Pr\{\pi_T(d_j) \geq \bar{\pi}_T  \,|\, \mathcal{D} \} \leq  c_T,   \\
{\rm  Acceptable\ Clinical\ Response\ Rate} & \hspace{2em}  \Pr\{\pi_R(d_j) \leq  \underline{\pi}_R  \,|\, \mathcal{D} \}  \leq  c_R,   \\
{\rm Acceptable\  Restricted\ Mean\ Survival\  Time} & \hspace{2em}   \Pr\{\mu_S(d_j) \leq  \underline{\mu}_S  \,|\, \mathcal{D} \} \leq c_S.
\end{align}
\label{eqn:accept3}
\end{subequations}
\noindent
These statistical requirements reflect the definition of OTD.   They say that,   given current data $\mathcal{D},$
to be an acceptable candidate for being selected as the OTD,   $d_j$ must be
 biologically active (2a), and unlikely to have  an unacceptably high toxicity probability  (2b),
   a low response probability (2c), or a
 short RMST (2d).   Under the model selection framework given above,  requiring $d_j\geq  \hat{\tau}_B$ is equivalent to requiring $\max_{d_{r}>d_{j}}\{\Pr(\mathcal{M}_{r} \,|\,\mathcal{D})\}\leq c_{B}$.
In each stage of the DEMO design,
some combination  of  the   four dose acceptability criteria  is applied, with each quantity  computed from the most recent data.
At the end of the trial, the estimated OTD using the final data $\mathcal{D}$ is  the dose in $\{d_1,\cdots, d_J\}$ that satisfies the acceptability conditions (2a) -- (2d) and
maximizes the posterior mean RMST, $E\{\mu_S(d_j)  \,|\, {\mathcal D}\}$, for $j=1,\cdots,J$.
The decision cutoffs  $c_B, c_T, c_R, c_S$ in (2a)  - (2d) are calibrated by computer simulation to obtain a design with good operating characteristics (OCs).

To compare the DEMO design to existing dose-finding designs,  we give conventional definitions of targeted doses, including the maximum tolerated dose (MTD) and optimal biological dose (OBD).  These definitions will be used in the simulation scenarios discussed in Section 6.   We will consider various scenarios, including cases where MTD $=$ OBD $=$ OTD, MTD $>$ OBD $=$ OTD, MTD $>$ OBD $\not =$OTD, and so on.
It  is important to clarify some logical inconsistencies with terminology used conventionally to define these target doses, however.  When a 3+3 algorithm is used in a phase l trial, the MTD  is defined statistically as the dose in  $\{d_1,\cdots, d_J\}$
selected by the 3+3 algorithm. That is, there is no optimality criterion used by a 3+3 algorithm, other than the algorithm itself.   If, instead,  a version of the CRM \citep{Cheung:2011book, yuan2022model} is used
in phase 1,
the MTD is defined as the dose for which the estimate of the true toxicity probability $\pi_T(d_j)$ is closest to a fixed value ${\pi}_T^{target}$, such as 0.25 or 0.30.

 It  is common practice to refer to the dose that maximizes  a phase 1-2 design's optimality criterion defined in terms of $(Y_R,Y_T)$ as an
 {\it optimal biological dose} (OBD),  despite the fact that no biological variable such as $Y_B$ is used.  This definition
 of OBD relies on an implicitly assumed surrogacy of $Y_R$ for biological activity, which is not true in general.
 This definition of OBD  actually identifies what more properly may be called an {\it optimal phase 1-2 dose}.
 In the present setting, we will consider phase 1-2 trials  run using an optimality criterion
 defined in terms of   a utility function $U(a, b)$
 that quantifies the desirability of each possible pair of outcomes $(Y_R=a, Y_T=b)$, for  $a,b=0,1.$
In practice, to establish $U(a,b)$,  it is convenient
to first set $U(1,0)$ = 100   for the best and  $U(0,1)$ = 0 for the worst possible outcomes, and then elicit the two intermediate values $U(0,0)$ and $U(1,1)$, possibly presenting the four numerical values of $U(a,b)$ in
a  $2 \times 2 $ utility table \citep{lin2020boin12,msaouel2023risk}.
The mean utility of dose $d_j$ is defined as $\bar{U}(d_{j})=E\{U(Y_{R},Y_{T}) \,|\, d_j\}$.
This construction may be generalized easily to accommodate bivariate ordinal outcomes \citep{thallpds}.
While this convention for defining an OBD is illogical in the present setting because  we observe and make explicit use of $Y_B,$ we will define OBD following this common practice.
Thus, given a utility, we define the OBD as the lowest dose  in $\{d_1,\cdots, d_J\}$ that is safe ($\pi_T(d_j)< \bar{\pi}_T$),
clinically effective ($\pi_R(d_j)>\underline{\pi}_R$), and maximizes $\bar{U}(d_{j})$.

 Our definition of the OTD  is far more stringent than the definitions of the MTD or OBD.
 By aiming for the OTD and
 assuming a  multivariate dose-outcome model for $p(Y_R, Y_T, Y_B, Y_S  \,|\, d)$
in which $Y_B$ may exert mediated effects on
each of the clinical variables $Y_R,$ $Y_T,$ and $Y_S$,  the DEMO design can be more effective in identifying a truly optimal dose.



\section{Stages of the DEMO Design}
\subsection{Design Overview}
\begin{figure}
 	\centering
 	\includegraphics[width=1\textwidth]{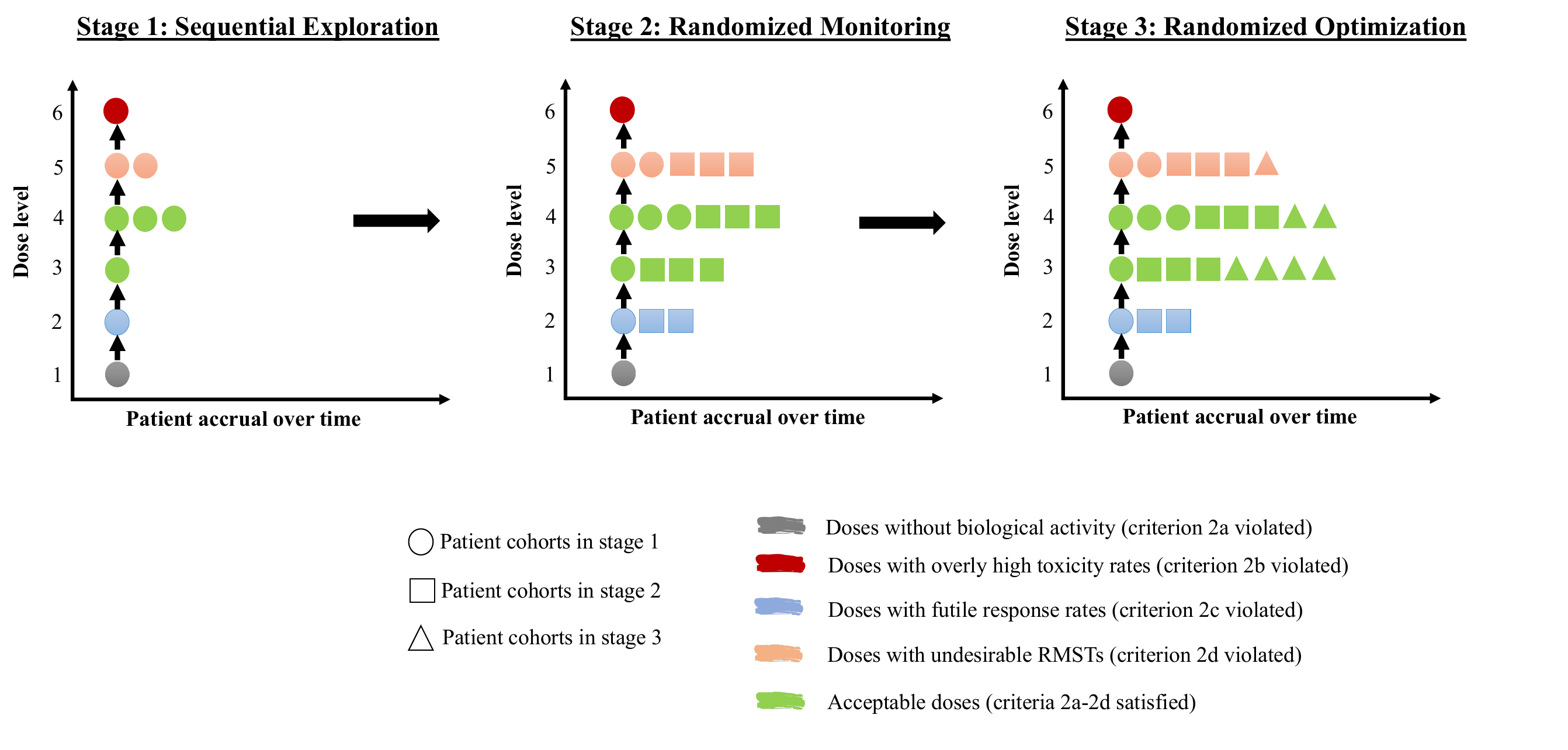}

\caption{Schematic of the three stages of the proposed dose exploration--monitoring--optimization (DEMO) design. Circles, rectangles, and triangles represent  patient cohorts enrolled in stages 1, 2, and 3, respectively.  The absence of rectangles or triangles at certain doses indicates that these doses do not meet at least one of the four acceptability criteria (2a)--(2b). RMST stands for the restricted mean survival time.  \label{fig: schema}}
\end{figure}

The DEMO design proceeds in  three stages,  assuming stage-specific models to evaluate the dose acceptability  criteria given above.
 Figure \ref{fig: schema} gives a schematic plot to illustrate the three stages, and additional details will be provided below in  subsequent sections.    If at any point of the trial there are no  admissibly safe, biologically active, and clinically effective doses, the trial is stopped and no dose is selected.
Stage 1  uses $Y_T,$ $Y_B,$ and possibly $Y_R$ to assign successive cohorts of patients to doses in sequentially adaptive steps,  while  eliminating doses that are unsafe or biologically inactive.
Stage 1 may use either a phase 1 or a phase 1-2 design's rules to assign acceptable doses to patient cohorts,
while also applying the biological acceptability rule (2a) and the safety rule (2b). In the example given in Figure \ref{fig: schema},  based on $Y_T$ and $Y_{B}$ collected during stage 1, the acceptable dose set  $\mathcal{A}_1=\{d_2,d_3,d_4,d_5\}$ is identified, since  $d_1$ is dropped due to  lack of bioactivity (criterion 2a) and $d_6$ is dropped due to excessive toxicity (criterion 2b).
In Stage 2, patients are randomized among acceptable doses, screening
for safety and biological activity is continued, and the design also screens doses for unacceptably low response rates (2c).
 In Figure \ref{fig: schema}, dose $d_2$  is eliminated in stage 2 due to insufficient efficacy, leading to  the acceptable dose set $\mathcal{A}_2= \{ d_3, d_4, d_5 \}$. At the end of stage 2,  if necessary to obtain a feasible design, the set of acceptable doses is further reduced to include at most $K$ doses, for $K\geq 2$.
Stage 3 continues to randomize patients among the acceptable doses, adds the RMST criterion (2d)  for restricting the set of acceptable doses, and chooses a final OTD based on RMST.  In  Figure \ref{fig: schema}, dose  $d_5$ is dropped due to  lack of sufficient survival benefit. At the end of stage 3, the RMST is estimated for each remaining acceptable dose based on the final data to determine the OTD.


\subsection{Stage 1: Sequential Exploration}
A maximum of $N_1$ patients are treated in Stage 1, applying the safety monitoring rule (2b),
 using either a phase 1 or phase 1-2 design  to choose acceptable doses for successive cohorts of patients.
 When $N_1/2$ patients have been treated and their biological variables $Y_B$ evaluated, the acceptability rule (2a) 
 is applied to drop biologically inactive doses.
 If  feasible, $Y_B$ may be  monitored more intensively,   such as when
  $N_1/3$ and $2N_1/3$ patients have been evaluated.

To apply the monitoring rules (2a) and (2b) feasibly in  stage 1, since the rules focus
on the marginal parameters,
we model $(Y_T,Y_B)$ data in stage 1 assuming a simple independence model, with $p(Y_T,Y_B|d)=p(Y_T \,|\, d)\times p(Y_B \,|\, d)$.  We assume that $p(Y_B \,|\, d)$ follows the above step function model, and that
toxicity follows a  Bayesian logistic regression model,   $\text{logit}\left\{ \pi_T(d_j ) \right\} = \alpha_0 + \alpha_1 d_j,$ where $\alpha_0$ is real-valued
 and $\alpha_1>0$. The  priors are assumed to be $\alpha_0\sim {\mathcal N}(m_{\alpha_0},\nu_{\alpha_0}^2)$ and $\log(\alpha_1) \sim {\mathcal N}(m_{\alpha_1},\nu_{\alpha_1}^2)$.  Elicitation of the hyperparameters $m_{\alpha_0},\nu_{\alpha_0}^2, m_{\alpha_1}$, and $\nu_{\alpha_1}^2$ is illustrated in the Supplementary Materials.



\subsection{Stage 2: Randomized Monitoring}
In stage 2, a maximum of $N_2$ additional patients are randomized among the  acceptable doses identified in stage 1,
applying the three acceptability criteria (2a), (2b), and (2c).
 Because in practice $Y_R$ may take a longer time to evaluate than $Y_T$ and $Y_B$,
the acceptability rule (2c) is added to take advantage of the fact that more response data will become available as the trial progresses.  Either simple or adaptive randomization   can be applied in  stage 2.

To monitor  dose acceptability while also
taking advantage of the biological effects of the agent when analyzing the
stage 2 data,   we assume that $Y_B$ is a mediator for $Y_R$ and $Y_T$.  We exploit  the  likelihood factorization
$
p(Y_R, Y_T, Y_B  \,|\, d) =     p(Y_R, Y_T  \,|\,  Y_B,  d) \times  p(Y_B  \,|\, d),
$
and  assume that
\begin{eqnarray}\label{jointmodel1}
\mu_{B}(d_{j}) &=& \gamma_{0}+\gamma_{1}d_{j}^{\gamma_{3}}/(\gamma_{2}^{\gamma_{3}}+d_{j}^{\gamma_{3}}),  \nonumber \\
\pi_{T}(Y_{B},d_{j}) &=& \Pr(Y_{T}=1 \,|\, Y_{B},d_{j})={\rm logit}^{-1}(\alpha_{0}+\alpha_{1}d_{j}+\alpha_{2}Y_{B}), \\
\pi_{R}(Y_{B},d_{j}) & =& \Pr(Y_{R}=1 \,|\, Y_{B},d_{j})={\rm logit}^{-1}(\beta_{0}+\beta_{1}d_{j}+\beta_{2}d_{j}^{2}+\beta_{3}Y_{B}). \nonumber
\end{eqnarray}
The  E$_{\max}$ model for $\mu_B(d)$ in expression (\ref{jointmodel1}), which now replaces the much simpler
step function  model  used in stage 1,
 is assumed for its flexibility.
The baseline parameter $\gamma_0$  is real-valued,
$\gamma_1>0$  is the maximum possible bioactivity,  $\gamma_2>0$  is the dose that produces half of $\gamma_1$, and $\gamma_3>0$  is the Hill factor that controls the steepness of $\mu_B(d)$.
The logistic regression model for $\pi_T$ as a function  of $d$ alone in stage 1   is elaborated
in stage 2 to include $Y_{B}$ as a covariate.
 We assume real-valued $\alpha_0$, with  $\alpha_1>0$ and  $\alpha_2 >0$ \
  to ensure that  $\pi_T(Y_{B},d_j)$ increases with both $d$ and $Y_B$.
  A similar model
$\pi_R(Y_{B},d_j)$ is assumed for $Y_R$, but
including a quadratic term to allow a more flexible dose effect.
  We do not make  monotonicity assumptions for $\pi_R(Y_{B},d_j)$, and allow
 $\beta_0, \ldots,\beta_3$ to take on any real values.

As tractable operational priors for the parameters in the joint model (\ref{jointmodel1}), we assume
 \begin{eqnarray*}
 &  & \gamma_{0}\sim\mathcal{N}(m_{\gamma},\nu_{\gamma}^{2}),\quad\gamma_{l}\overset{\rm i.i.d.}{\sim}{\rm Gamma}(a_{\gamma_{l}},b_{\gamma_{l}}),l=1,2,3,\quad\sigma_{ B }^{-2}\sim{\rm Gamma}(a_{\sigma},b_{\sigma}),\\
 &  & \log\alpha_{l}\overset{\rm i.i.d.}{\sim}\mathcal{N}(m_{\alpha_{l}},\nu_{\alpha_{l}}^{2}),l=1,2,\quad\beta_{l}\overset{\rm i.i.d.}{\sim}\mathcal{N}(m_{\beta_{l}},\nu_{\beta_{l}}^{2}),l=1,2,3,  \\
 & & (\alpha_{0},\beta_{0})\sim\mathcal{BN}\left((\mu_{\alpha_{0}},\mu_{\beta_{0}}),(\nu_{\alpha_{0}}^{2},\nu_{\beta_{0}}^{2},\rho_{0})\right),
 \end{eqnarray*}
where $\tilde{\btheta}$ = $(m_{\gamma},\nu_{\gamma}^{2}$, $a_{\gamma_{l}},b_{\gamma_{l}}$, $a_{\sigma},b_{\sigma}$, $m_{\alpha_{l}},\nu_{\alpha_{l}}^{2}$, $m_{\beta_{l}},\nu_{\beta_{l}}^{2}$, $\mu_{\alpha_{0}}, \mu_{\beta_{0}}$,
$\nu_{\alpha_{0}}^{2}$, $\nu_{\beta_{0}}^{2}$,$\rho_0)$ is the vector of all model  hyperparameters, and $\mathcal{BN}$ denotes the bivariate normal distribution with correlation $\rho_0$.
 Guidelines for specifying $\tilde{\btheta}$ 
 are given in the Supplementary Materials.
  We use a Gibbs sampler to obtain posterior samples of the parameters in the joint model (\ref{jointmodel1}).
  To avoid numerical integration of $\pi_T(Y_B,d_j)$ and $\pi_R(Y_B,d_j)$
  over the distribution of $Y_B$ when computing the marginals $\pi_T(d_j)$ and
  $\pi_R(d_j)$, we use a plug-in approach by substituting the posterior mean
  $\hat{\mu}_{B}(d_j)$ in place of $Y_B$ to approximate $\pi_T(d_j)$ and $\pi_R(d_j)$ for each $d_j.$
  This facilitates computing the acceptability rules (2a)--(2c).

\subsection{Stage 3: Randomized Optimization}

  In stage 3, a maximum of $N_3$ additional patients are randomized among the current set of  acceptable doses,
applying the four acceptability criteria (2a), (2b), (2c), and (2d).
If at any point there are no  acceptable doses, the trial is stopped and no dose is selected.
Otherwise, the dose with largest posterior mean RMST based on the final data  is selected as the OTD.

The set of doses to be studied in stage 3 is determined as follows.
  First, for a small fixed integer $L\geq 2$, adaptively select the best $L$ acceptable doses, denoted by $\mathcal{C}_1$, in terms of  the posterior mean response rate,
  or the mean utility.
To account for plateau scenarios  where multiple doses have similar response rates,
 we define set $\mathcal{C}_2$ to be the best $K$ ($K\geq L$) acceptable doses with posterior tail probabilities of $\pi_R(d_j)$ at least a specified fraction $(1-\kappa) \in (0, 1)$ of the maximum value, that is,
\[
\Pr\{\pi_{R}(d)>\underline{\pi}_{R}  \,|\, {\mathcal D} \} \geq (1-\kappa)\max_{d}\{\Pr(\pi_{R}(d)>\underline{\pi}_{R}   \,|\, {\mathcal D} ) \}.
\]The set of   doses selected for stage 3 evaluation  then is defined as
$ \mathcal{C}_{1}\cup\mathcal{C}_{2},$  and the number of doses is denoted by $W=|\mathcal{C}_{1}\cup\mathcal{C}_{2}|$.    Intuitively, suppose that the top $L$  acceptable doses are advanced to the final stage. If there are many promising doses demonstrating nearly equivalent maximum response rates,  at most an additional $K-L$ doses will be added for stage 3. When $\kappa=0$, this gives  a smaller set with $ \leq L$ doses, while $\kappa=1$ gives a larger set with $\leq K$ doses.

 To obtain a feasible sample size, depending on the number of doses $J$ specified at the beginning of the trial, as a heuristic rule we suggest using $(L,K,\kappa)=(2,3,0.3)$ for $J\leq 5$ and  $(L,K,\kappa)=(3,4,0.3)$ for $J\geq 6$.   If $W$ is impractically large, one can further reduce $L$ and $K$ to obtain
a manageable stage 3  dose set.
 It  also is important to note that the number of selected doses $W$ may be smaller than $L$ due to the three acceptability rules (2a)--(2c) imposed through stages 1 and 2.
According to our sensitivity analyses  reported below, in Section \ref{simulaiton-study},   the  design yields robust performance using the default specifications of $(L,K,\kappa)$.

In stage 3,  patients are  randomized unequally  among the  selected doses, restricted  to ensure that each  dose has a maximum sample size of $M$. As an illustration, if $M$ = 20, and the $W$ = 3 selected doses have respective sample sizes  12, 9, 14 at the start of stage 3, then the dose-specific sample sizes in stage 3 would be $M-n_{1} = 8, M-n_{2} = 11,$ and  $M-n_{3}$ = 6. Another randomization strategy involves using response-adaptive randomization, which allocates more patients to the better-performing arms, while capping the maximum sample size at a reasonable value.

To evaluate the dose acceptability rules (2a)--2(d), we construct the stage 3-specific full model using the likelihood factorization
\[
p(Y_S, Y_R, Y_T, Y_B  \,|\, d) =   p(Y_S  \,|\, Y_R, Y_T, Y_B,  d)  \times  p(Y_R, Y_T  \,|\,  Y_B,  d) \times  p(Y_B  \,|\, d).
\]
Building on the stage 2-specific joint model for $ p(Y_R, Y_T  \,|\,  Y_B,  d) \times  p(Y_B  \,|\, d)$, we  assume that $p(Y_S  \,|\, Y_R, Y_T, Y_B,  d)$ is a Weibull regression model  with hazard function
\begin{equation}\label{weibulls}
h(Y_S  \,|\, d_j, Y_T, Y_R,Y_B) =  \rho Y_S^{\rho-1} \lambda_{j} \exp \left\{\eta_1 Y_T + \eta_2 Y_R +\eta_3 Y_B \right\}
\end{equation}
where $\rho\geq 1$ is the shape parameter, and each $\lambda_{j}>0$ is a dose-dependent scale parameter that characterizes the direct effect of dose $d_j$ on survival.  The real-valued parameters  $\eta_1,\eta_2, \eta_3$  quantify  indirect dose effects on survival mediated through $Y_T$, $Y_R$ and $Y_B$, respectively (see Figure \ref{fig:DAG}). For priors of the parameters in (\ref{weibulls}), we assume   $\rho\sim {\rm Gamma}(a_\rho,b_\rho)$, $\log(\lambda_j)\sim \mathcal{N}(0, \nu_\lambda^2)$, and $\eta_1,\eta_2,\eta_3 \overset{\rm i.i.d.}{\sim} \mathcal{N}(0, \nu_\eta^2)$, where $a_\rho$, $b_\rho$, $\nu_\lambda^2$, and $\nu_\eta^2$ are hyperparameters 
included in $\tilde{\btheta}$, with suggested values provided in the Supplementary Materials.




Based on the final data $\mathcal{D}$, we jointly estimate  models (\ref{jointmodel1}) and (\ref{weibulls}) using  a Gibbs sampler. To obtain posterior estimates of RMST,  at each sampling iteration, we first compute the posterior mean of ${\mu}_{B}(d_j)$ under the E$_{\max}$ model, denoted by $\hat{\mu}_{B}(d_j)$, and the posterior mean of the joint probability of toxicity and response at dose $d_j$, denoted by $ \hat{\pi}_j(u, v),$ for $u,v=0,1.$
The RMST
over a fixed  period $t_S$ at dose $d_j$ then can be estimated by the plug-in method,
$$\hat{\mu}_S(d_j)=   \sum_{u=0}^1 \sum_{v=0}^{1} \hat{\pi}_j(u, v) \int_0^{t_S} \exp\{-\hat{\lambda}_{j}Y_{S}^{\hat{\rho}}\exp(\hat{\eta}_{1}u+\hat{\eta}_{2}v+\hat{\eta}_{3}\hat{\mu}_B(d_j))\}{\rm d}Y_{S}.$$
The OTD   then is selected as the dose that satisfies (2a)--(2d) and maximizes  $ \hat{\mu}_S(d_j)$.


\section{Illustration}
\label{trialillustration}
To illustrate DEMO,  we  apply it to redesign and conduct hypothetical versions of the DREAMM-1 and DREAMM-2 studies,  investigating  six doses, $(d_1,\ldots,d_6)$ =(0.48, 0.96, 1.92, 2.5, 3.4, 4.5) mg/kg.
For each patient, we simulate  data using the estimated event rates and survival times based on the published  trial data. This gives  estimated  $\pi_R(d)$ rates   $(0.06,\, 0.11,\, 0.18,\, 0.31,\, 0.33,\, 0.34)$. Since no DLT was observed in DREAMM-1, we  assume that $\pi_T(d)$  rates are the negligible values $(0.01,\, 0.02,\, 0.03,\, 0.04,\, 0.05, \, 0.06)$.
To generate the biological outcome $Y_B$,  which is the logarithmic change in BCMA concentration,
we use equation (\ref{Bmodel}) assuming $(\mu_{B}(d_1),\ldots,\mu_{B}(d_6))$ =(3.88, 5.50, 5.93, 5.97, 5.99, 6.00) and $\sigma_{B}^2=1$, which gives  $d_1$ as biologically inactive, and $\{d_2, \cdots, d_6\}$ as active.
The published results  only provide median PFS times for doses $d_4$ and $d_5$, given by 11.0 and 13.7 months, respectively.
We thus assume that the median PFS times for the remaining doses $d_1, d_2, d_3$, and $d_6$, are 5.7, 6.5, 6.9, and 10.2 months, respectively. Based on the specified median  times, we simulated the PFS time from the Weibull regression model (\ref{weibulls}) by fixing $(\lambda_1,\ldots,\lambda_6)=(0.11,\, 0.10,\, 0.10,\, 0.07,\, 0.06, \, 0.08)$, $\rho=1.1$, $\eta_1=3$, and $\eta_2=-2$. Each patient was  followed up to 36 months, and only administrative censoring was considered. The underlying RMSTs over two years for the six doses were then 7.7, 8.7, 9.5, 12.8, 13.7, and 12.2 months, respectively. To implement the DEMO design, we set the limits  $\bar{\pi}_T=0.25$, $\underline{\pi}_R=0.15$, and $\underline{\mu}_S=9$, and, based on preliminary simulations.  specified acceptability cutoffs $c_{B}=0.5$, $c_T=0.6$, $c_R=0.7$, and $c_S=0.8$.  We also adopted the default specification of the hyperparameters as described in the Supplementary Materials, by setting  $(L,K,\kappa)=(3,4,0.3)$. In this hypothetical setting, $d_5$ was the OTD.

 Summary data at each decision-making time for the hypothetical trial are given in Table \ref{trial_illustration}. In  stage 1, $27$ patients were  allocated among the six doses using the  BOIN design with a cohort size of three. Based on the $Y_T$ data, all doses satisfied (2a). Analysis of the $Y_B$  data showed that doses no lower than  $\hat{\tau}_{B}=2$ were likely to exhibit promising bioactivity, and the acceptable dose set  at the end of  stage 1 was $(d_2,d_3,d_4,d_5,d_6)$. In stage 2,  three new cohorts of patients were assigned to each acceptable dose, with three patients in each cohort.  Dose $d_2$ was eliminated by the clinical response rate futility criterion
  (2c), so the acceptable dose set after stage 2 was   $(d_3,d_4,d_5,d_6)$.

In stage 3,  doses $d_4$, $d_5$, and $d_6$ were selected to treat further patients, because  $d_3$ falied to satisfy the selection rule, with   $\Pr(\pi_R(d_3) > 0.15  \,|\, \mathcal{D})  < 0.7 \Pr(\pi_R(d_6) > 0.15  \,|\, \mathcal{D}) $. In the final stage,  no further dose elimination was triggered, and 12, 12, and 6 additional  patients were randomized to $d_4$, $d_5$, and $d_6$  to achieve a total per-dose sample size of 24 patients.   After  completion of the trial, the estimated two-year RMST values were 13.9, 14.9, and 12.1 months for  $d_4$, $d_5$ and $d_6$, respectively. Consequently, based on the acceptability criteria (2a)--(2d) and the final estimated RMST values,
 the DEMO design identified the OTD to be $d_5$ (3.4 mg/kg).



\section{Simulation Study}
\label{simulaiton-study}
\subsection{Simulation Setting}
We conducted a simulation study to evaluate the performance of the DEMO design, considering  six doses $(d_1, \ldots, d_6) = (0.05, 0.10, 0.20, 0.45, 0.65, 0.85)$.   For simulated trials, during  stage 1, a maximum of 10 cohorts of size three patients each were assigned to doses using the BOIN design with target toxicity rate  ${\pi}_T^{target}=0.30$ and the default parameter settings for  the BOIN design.  Stage 1  would terminate if the maximum sample size of 30 patients was reached or if a maximum of nine patients had been treated at a particular dose, whichever came first. In stage 2, a maximum of nine patients were assigned to each acceptable dose, with interim monitoring rules applied after every three patients.
Stage 3 randomized patients until each acceptable dose reached a maximum of 24 patients.  One interim analysis was performed midway through stage 3.
We specified $\bar{\pi}_T = 0.30$, $\underline{\pi}_E = 0.20$, and $\underline{\mu}_S = 3$.
Based on  preliminary simulations, the probability cutoffs for acceptability rules (2a)--(2d) were calibrated to be  $c_T=0.60$, $c_B=0.50$, $c_R=0.70$, and $c_S=0.80$.
 Prior hyperparameter values  of the DEMO design are provided in the Supplementary Materials, and we set $(L, K, \kappa) = (3, 4, 0.3)$.

 We included two other designs as comparators.
 The first comparator, referred to as DFCE, utilized a dose finding  plus randomized cohort expansion strategy by employing BOIN to estimate the MTD during the dose-finding stage,  followed by  expansion of the three dose levels MTD, MTD-1, and MTD-2,  with a maximum sample size of 24 at each dose. The DFCE design uses the same BOIN configuration and trial monitoring procedures as DEMO, and selects the acceptable dose with highest RMST as the OTD. The  two differences between DFCE and DEMO are that  DFCE does not utilize the  endpoint $Y_B$ (i.e., the acceptability rule (2a)) to determine an acceptable dose, and the selection of randomized doses by DFCE  depends solely on $Y_T$ and the  estimated MTD.

The second comparator was the two-stage utility-based U-BOIN design \citep{zhou2019utility}. Similarly to DEMO, the first stage of U-BOIN utilizes BOIN to explore the dose space, which terminates when a particular dose has accumulated  nine patients. and then identifies acceptable doses  based on binary $Y_T$ and $Y_R$ data. In the second stage, U-BOIN randomizes patients equally  among acceptable doses until the maximum sample size of 120 is reached, or if the number of patients treated at any dose reaches 30.   The utility values for the joint outcomes $(Y_R, Y_T) = \{ (0, 1), (0, 0), (1, 1), (1, 0) \}$ were set to  (0, 5, 95, 100).
The results of U-BOIN were obtained using the software available on www.trialdesign.org, without changing any other design parameters. The key difference between U-BOIN and DEMO is that U-BOIN  only considers binary $Y_T$ and $Y_R$ data for identifying the OBD, while DEMO incorporates additional $Y_B$ and $Y_S$ data to determine the OTD.



Ten  scenarios were considered, with varying configurations of underlying assumed true $Y_B$ means, toxicity and response rates, and one-year RMST values (Table \ref{tab:scenario}). The $Y_B$ data were generated from   (1) with variance $\sigma_{B}^2=1$. The $Y_S$ data were generated from (\ref{weibulls}) by fixing   $\rho=1.5$, $\eta_1=3$,  $\eta_2=-2$, and a two-year trial duration.  The parameters $\lambda_1,\ldots,\lambda_J$ were obtained by back-solving
to ensure that the specified RMST values could be achieved.  In scenarios 1 and 4, the three target doses (i.e., MTD, OBD, and OTD) align at the same dose level. In scenarios 2 and 3, the MTD and OBD are at the same dose level and higher than the OTD. In scenarios 5, 8, and 9, the OTD coincides with the OBD, but they are lower than the MTD. In scenarios 6 and 7, all three target doses are at different levels. In scenario 10, all the six dose levels are either futile or overly toxic.

\subsection{Simulation Results}
A total of 1000 trials were simulated for each scenario and design. The results are presented in Table \ref{tab:PCS}. Overall, across the ten scenarios, the DEMO design consistently gives superior performance in terms of the percentage of correct OTD selections (PCS)  compared with both the DFCE and U-BOIN designs.   DEMO  outperforms  U-BOIN primarily due to its utilization of $Y_S$  as one of the endpoints.
Compared to DFCE, DEMO exhibits  superior performance  in most scenarios when the OTD is not one of the doses  MTD, MTD-1, and MTD-2, such as in scenario 9. Even when the OTD equals the MTD,  as in scenario 1,  DEMO also outperforms DFCE, because using $Y_B$ in later stages improves efficiency.

Table \ref{tab:PCS} also summarizes the average number of patients required for each design in each scenario.
Because of its termination rule, U-BOIN  has  the smallest sample sizes in scenarios 1 to 9. In scenario 10, where no doses are acceptable,  DEMO   correctly terminates all doses and leads to a smaller sample size than  U-BOIN.
The results also show that  DEMO   requires fewer patients than DFCE in most scenarios, mainly because  incorporation of early $Y_B$ data helps to reduce the acceptable dose set.   Among the scenarios where DEMO requires more patients, a typical example is scenario 9. This is due to DEMO's greater flexibility in selecting stage 3 doses, in contrast with DFCE, which is limited to only three doses. Consequently, despite the larger sample size required by DEMO in scenario 9, a substantial improvement of 77.9\% in the PCS is observed compared to DFCE.

\subsection{Sensitivity Analysis}
In the  Supplementary Materials, we provide sensitivity analysis results to demonstrate the robustness of the DEMO design across various prior specifications. 
We also report an additional simulation study to evaluate design OCs for different values of $(L,K,\kappa)$.

\section{Discussion}

In this paper, we have proposed a novel three-stage seamless generalized  phase 1-2 clinical trial design for identifying an optimal dose that includes information from a biological variable, toxicity, response, and survival time.
The  DEMO design is structured to follow the actual process of outcome evaluation
and decision making in an early-phase dose-finding trial, where the data are sparse at the beginning of the trial and become richer as the trial proceeds.   The statistical model and decision-making process underlying DEMO are elaborated during successive stages of the trial to make use of accumulating  data.

The simulations showed that the  DEMO design is  superior to traditional designs, in terms of  both correct decision percentages and sample sizes,  while treating a greater proportion of subjects at the optimal dose.
Incorporating a biological variable provided  greater efficiency for both screening out ineffective doses 
and choosing an OTD. 
Sensitivity analyses  confirmed that the design is robust to changes in  prior specifications.   Selection of the design parameters $(L,K,\kappa)$ may reflect  causal relationships between the bioactivity, toxicity, response, and survival time. While  smaller values for $(L,K,\kappa)$ may be appropriate if there are strong causal relationships with long-term survival {\it a priori}. Since, in practice, such an association is uncertain,  we recommend setting $(L,K,\kappa)=(2,3,0.3)$ for trials with $J\leq 5$ doses and  $(L,K,\kappa)=(3,4,0.3)$ for trials with $J\geq 6$ doses, respectively. Further  generalizations of DEMO may include incorporating multiple biological endpoints,
longitudinally observed biomarker processes, and optimizing subgroup-specific doses for  trials that enroll patients with different tumor types.

\bibliographystyle{abbrvnat} 
\bibliography{biomsample-v6}

\begin{table}[h]
{
\caption{Data summary, including number of patients, sample mean of biological variable $Y_B$, number of toxicities $Y_T$, number of  responses $Y_R$, 24-month restricted mean survival time (RMST),  median progression-free survival time (PFS, in months), and decision-making statistics for the hypothetical trial described in Section \ref{trialillustration}. } \label{trial_illustration}
\centering
\def\arraystretch{1.25}%
\begin{center}
\begin{tabular}{lllllllllllllll}
\hline
                    &    & \multicolumn{6}{c}{Dose level}              &          & \multicolumn{6}{|c}{Dose level}               \\
                    &    & 1        & 2      & 3         & 4       & 5      & 6   &    & 1      & 2       & 3     & 4     & 5    & 6        \\
                \hline
                     &   & \multicolumn{6}{c}{End of Stage 1}                 &               & \multicolumn{6}{c}{End of Stage 2}                        \\
Number of patients   &         & 3      & 3      & 3     & 3     & 3     & 9     &           & 3      & 12     & 12    & 12    & 12   & 18        \\
Sample mean of $Y_B$   &       & 4.1    & 6.1    & 5.7   & 5.6   & 5.2   & 5.6     &         & 4.1    & 5.3    & 5.6   & 5.8   & 6.1  & 5.6   \\
Number of $Y_T$   &          & 0    & 0   & 0   & 0  & 0   & 0         &     & 0     & 0     & 0    & 1     & 0   & 1      \\
Number of $Y_R$    &        & 0    & 0    & 2   & 1   & 1   & 3         &    & 0    & 1      & 2     & 3     & 3    & 6       \\
RMST        &          & 2.5    & 1.6    & 19.4   & 8.4   & 15.6  & 14.9        &      & 2.5    & 9.4    & 9.0   & 11.0   & 15.3  & 14.2    \\
Median PFS  &  & 3.1    & 1.9    & NA   & 6.3   & 18.7  & 13.4          &    & 3.1    & 6.9    & 6.7   & 8.2   & 16.2  & 14.0   \\
$\Pr(\mathcal{M}_j|\mathcal{D})$     &         & 0.00     & 0.89     & 0.04    & 0.03    & 0.02   & 0.02 &   & 0.00   & 0.84   & 0.11   & 0.04   & 0.01   & 0.00         \\
$\Pr({\pi}_T(d_j) > 0.25|\mathcal{D})$ && 0.00     & 0.00     & 0.00    & 0.00    & 0.02   & 0.16  &  & 0.00   & 0.00   & 0.00   & 0.00   & 0.00  & 0.00      \\
$\Pr({\pi}_E(d_j) < 0.15|\mathcal{D})$& & NA       & NA       & NA      & NA      & NA     & NA   &   & 0.86   & 0.85   & 0.48   & 0.22   & 0.04   & 0.03        \\
acceptable?    &                  & No        & Yes        & Yes       & Yes       & Yes      & Yes   &    & No      & No      & Yes      & Yes      & Yes      & Yes           \\
\hline
                    &    & \multicolumn{6}{c}{1st interim of Stage 2}          &     & \multicolumn{6}{c}{1st interim of Stage 3}       \\
Number of patients    &         & 3      & 6      & 6     & 6     & 6     & 12     &          & 3      & 12     & 12    & 18    & 18    & 21    \\
Sample mean of $Y_B$   &      & 4.1    & 5.9    & 5.6   & 6.0   & 6.0   & 5.7       &       & 4.1    & 5.3    & 5.6   & 5.8   & 5.9   & 5.7   \\
Number of $Y_T$    &         & 0      & 0      & 0     & 0     & 0     & 0       &        & 0      & 0      & 0     & 1     & 0     & 1          \\
Number of $Y_R$      &         & 0      & 0      & 2    & 1     & 2     & 4       &         & 0      & 1      & 2     & 4     & 5     & 6      \\
RMST           &        & 2.5    & 7.9    & 11.8   & 7.9   & 13.5  & 15.1       &       & 2.5    & 9.4    & 9.0   & 13.0   & 14.5   & 12.7       \\
Median PFS  &  & 3.1   & 4.3    & 8.3   & 6.4   & 12.4  & 14.3        &      & 3.1    & 6.9    & 6.7   & 10.1   & 14.9  & 12.8  \\
$\Pr(\mathcal{M}_j|\mathcal{D})$     &         & 0.00     & 0.95     & 0.02    & 0.01    & 0.01   & 0.01  &  & 0.00     & 0.83    & 0.12     & 0.04     & 0.01     & 0.00       \\
$\Pr({\pi}_T(d_j) > 0.25|\mathcal{D})$& & 0.00     & 0.00     & 0.00    & 0.00    & 0.00   & 0.00  &  & 0.00   & 0.00   & 0.00   & 0.00   & 0.00   & 0.00    \\
$\Pr({\pi}_E(d_j) < 0.15|\mathcal{D})$& & NA       & NA       & NA      & NA      & NA     & NA   &   & 0.92   & 0.88   & 0.48   & 0.17   & 0.02   & 0.05     \\
acceptable?    &                  & No        & Yes        & Yes       & Yes       & Yes      & Yes   &    & No      & No      & Yes      & Yes      & Yes      & Yes           \\
\hline
                   &     & \multicolumn{6}{c}{2nd interim of Stage 2}     &         & \multicolumn{6}{c}{End of Stage 3}               \\
Number of patients   &           & 3      & 9      & 9     & 9     & 9     & 15   &            & 3      & 12     & 12    & 24    & 24    & 24      \\
Sample mean of $Y_B$   &       & 4.1    & 5.5    & 5.5   & 6.0  & 6.2   & 5.5      &        & 4.1    & 5.8    & 6.0   & 6.2   & 5.8   & 6.2      \\
Number of $Y_T$    &         & 0      & 0      & 0     & 0    & 0    & 0      &          & 0      & 0     & 0    & 1     & 0    & 1         \\
Number of $Y_R$     &           & 0      & 1      & 2     & 3     & 3    & 5     &           & 0      & 1      & 2     & 6    & 7    & 7         \\
RMST        &           & 2.5    & 10.0    & 10.5   & 10.8   & 15.8   & 15.4        &      & 2.5    & 9.4    & 9.0   & 12.9   & 13.9   & 12.5     \\
Median PFS  &  & 3.1    & 7.4    & 7.5   & 6.5   & 18.7  & 15.3   &           & 3.1    & 6.9    & 6.7   & 9.9   & 13.6  & 11.9   \\
$\Pr(\mathcal{M}_j|\mathcal{D})$     &    & 0.00     & 0.92     & 0.04    & 0.03    & 0.01   & 0.00  &  & 0.00     & 0.81     & 0.14    & 0.05     & 0.00     & 0.00          \\
$\Pr({\pi}_T(d_j) > 0.25|\mathcal{D})$ && 0.00     & 0.00     & 0.00    & 0.00    & 0.00   & 0.00  & & 0.00   & 0.00   & 0.00   & 0.00   & 0.00   & 0.00    \\
$\Pr({\pi}_E(d_j) < 0.15|\mathcal{D})$& & NA       & NA       & NA      & NA      & NA     & NA  &    & 0.93   & 0.88   & 0.36  & 0.08   & 0.05   & 0.04   \\
acceptable?      &                & No        & Yes        & Yes       & Yes       & Yes      & Yes    &   & No      & No      & Yes      & Yes      & Yes      & Yes        \\
\hline
\hline
\end{tabular}
\end{center}
}
\vspace{0.35in}

{\footnotesize Note: ``NA'' means either the data were not available or the median survival time was not reached.  Reported RMST  values were obtained based on the estimated Kaplan-Meier survival curve. At the end of stage 3, the RMSTs estimated under the proposed Weibull regression model were (3.7, 9.8, 9.3, 13.9, 14.9, 12.1) months.}
\end{table}

\begin{table}[h]
\renewcommand{\arraystretch}{1.4}
{
\caption{Assumed true means of the biological outcome $\mu_B(d)$, toxicity probabilities $\pi_T(d)$, response probabilities $\pi_R(d)$, and one-year restricted mean survival times $\mu_S(d)$ for each dose under each
simulation scenario. The optimal therapeutic dose (OTD) is given in boldface.
}
\label{tab:scenario}
  \begin{center}
\begin{tabular}{lllllllllllll}
\hline
Dose level & 1    & 2    & 3    & 4    & 5    & 6 \hspace{1cm}   & 1    & 2    & 3    & 4    & 5    & 6    \\
\hline
           & \multicolumn{6}{c}{scenario 1}                      & \multicolumn{6}{c}{scenario 6}          \\
$\mu_B(d)$         & 2.00 & 2.01 & 2.08 & 2.76 & 3.75 & \textbf{4.73} & 2.24 & 4.00 & \textbf{5.77} & 5.99 & 6.00 & 6.00 \\
$\pi_T(d)$   & 0.01 & 0.02 & 0.03 & 0.06 & 0.13 & \textbf{0.26}    & 0.01 & 0.02 & \textbf{0.05} & 0.10 & 0.27 & 0.55 \\
$\pi_R(d)$    & 0.04 & 0.05 & 0.08 & 0.20 & 0.35 & \textbf{0.47}    & 0.07 & 0.14 & \textbf{0.32} & 0.41 & 0.42 & 0.44 \\
$\mu_S(d)$        & 1.15 & 1.42 & 1.51 & 3.14 & 4.04 & \textbf{5.35}   & 1.95 & 4.86 & \textbf{5.70} & 3.66 & 3.12 & 2.20 \\
           & \multicolumn{6}{c}{scenario 2}                      & \multicolumn{6}{c}{scenario 7}                    \\
$\mu_B(d)$         & 2.01 & 2.09 & 2.24 & 4.17 & \textbf{5.29} & 5.95 & 5.04 & \textbf{5.83} & 5.98 & 6.00 & 6.00 & 6.00  \\
$\pi_T(d)$     & 0.01 & 0.03 & 0.04 & 0.07 & \textbf{0.14} & 0.28    & 0.01 & \textbf{0.06} & 0.18 & 0.29 & 0.51 & 0.54  \\
$\pi_R(d)$   & 0.04 & 0.06 & 0.09 & 0.23 & \textbf{0.37} & 0.44    & 0.22 & \textbf{0.35} & 0.41 & 0.42 & 0.44 & 0.45  \\
$\mu_S(d)$       & 1.15 & 1.87 & 2.37 & 2.94 & \textbf{4.10} & 2.91   & 4.90 & \textbf{5.97} & 4.13 & 3.05 & 2.35 & 2.27  \\
           & \multicolumn{6}{c}{scenario 3}                      & \multicolumn{6}{c}{scenario 8}          \\
$\mu_B(d)$        & 2.02 & 2.18 & 3.14 & \textbf{5.86} & 6.55 & 6.79 & 3.50 & \textbf{5.71} & \textbf{5.98} & 5.99 & 6.00 & 6.00 \\
$\pi_T(d)$    & 0.03 & 0.04 & 0.07 & \textbf{0.09} & 0.15 & 0.29    & 0.01 & \textbf{0.10} & \textbf{0.11} & 0.25 & 0.45 & 0.56 \\
$\pi_R(d)$  & 0.04 & 0.08 & 0.17 & \textbf{0.38} & 0.46 & 0.46    & 0.15 & \textbf{0.42} & \textbf{0.43} & 0.46 & 0.48 & 0.50 \\
$\mu_S(d)$      & 1.13 & 1.94 & 2.67 & \textbf{4.35} & 3.18 & 2.74   & 3.86 & \textbf{6.28} & \textbf{6.33} & 4.32 & 3.50 & 3.13 \\
           & \multicolumn{6}{c}{scenario 4}                      & \multicolumn{6}{c}{scenario 9}                   \\
$\mu_B(d)$       & 2.00 & 2.01 & 2.13 & 3.98 & \textbf{5.70} & 6.47 & 3.50 & 5.85 & \textbf{5.99} & 6.00 & 6.00 & 6.00 \\
$\pi_T(d)$    & 0.01 & 0.02 & 0.04 & 0.08 & \textbf{0.20} & 0.56    & 0.01 & 0.02 & \textbf{0.03} & 0.04 & 0.05 & 0.06 \\
$\pi_R(d)$   & 0.04 & 0.05 & 0.10 & 0.30 & \textbf{0.43} & 0.44    & 0.08 & 0.24 & \textbf{0.42} & 0.43 & 0.43 & 0.43 \\
$\mu_S(d)$     & 1.40 & 1.85 & 2.41 & 3.98 & \textbf{5.71} & 5.65   & 2.38 & 3.88 & \textbf{6.34} & 4.81 & 4.77 & 4.73 \\
           & \multicolumn{6}{c}{scenario 5}                      & \multicolumn{6}{c}{scenario 10}      \\
$\mu_B(d)$        & 2.24 & 2.80 & 4.00 & \textbf{5.34} & 5.65 & 5.79 & 2.90 & 4.25 & 5.60 & 6.29 & 6.40 & 6.44 \\
$\pi_T(d)$    & 0.01 & 0.02 & 0.04 & \textbf{0.08} & 0.26 & 0.52    & 0.32 & 0.36 & 0.41 & 0.42 & 0.48 & 0.52 \\
$\pi_R(d)$   & 0.04 & 0.06 & 0.14 & \textbf{0.40} & 0.38 & 0.33    & 0.02 & 0.03 & 0.10 & 0.12 & 0.13 & 0.17 \\
$\mu_S(d)$        & 1.40 & 1.87 & 3.26 & \textbf{5.97} & 3.67 & 2.49   & 1.99 & 2.07 & 2.36 & 2.59 & 2.63 & 2.89 \\
\hline
\end{tabular}
\end{center}
}
\end{table}

\begin{table}[!h]
\caption{Selection percentages (Sel\%) and number of patients (Pts) at each dose   for each design under scenarios 1 to 10  given in Table 2. The optimal therapeutical dose (OTD) under each scenario is given in boldface. ``None'' is the percentage of trials without selecting any dose, and ``N''  is the average total sample size. DEMO is the proposed design for dose exploration, monitoring, optimization; DFCE is the design that integrates dose finding (DF) with  cohort expansion (CE) strategies; and U-BOIN is the utility-based Bayesian optimal interval design.}
\label{tab:PCS}
\centering\begingroup\fontsize{8.5}{12}\selectfont
\setlength{\tabcolsep}{4pt}
  \begin{center}
\begin{tabular}[t]{lrrrrrrlrrrrrr}
\toprule
\multicolumn{1}{c}{} & \multicolumn{2}{c}{DEMO} & \multicolumn{2}{c}{DFCE} & \multicolumn{2}{c}{U-BOIN} & \multicolumn{1}{c}{} & \multicolumn{2}{c}{DEMO} & \multicolumn{2}{c}{DFCE} & \multicolumn{2}{c}{U-BOIN} \\
\cmidrule(l{3pt}r{3pt}){2-3} \cmidrule(l{3pt}r{3pt}){4-5} \cmidrule(l{3pt}r{3pt}){6-7} \cmidrule(l{3pt}r{3pt}){9-10} \cmidrule(l{3pt}r{3pt}){11-12} \cmidrule(l{3pt}r{3pt}){13-14}
Dose & Sel\% & Pts & Sel\% & Pts & Sel\% & Pts &  & Sel\% & Pts & Sel\% & Pts & Sel\% & Pts  \\
\midrule
\multicolumn{1}{c}{} & \multicolumn{6}{c}{Scenario 1} & \multicolumn{1}{c}{}  & \multicolumn{6}{c}{Scenario 6}   \\
\hspace{1em}1 & 0.0 & 3.8 & 0.0 & 3.4 & 0.4 & 3.3 & & 0.0 & 3.2 & 0.1 & 3.6 & 0.5 & 3.6\\
\hspace{1em}2 & 0.0 & 4.0 & 0.0 & 3.9 & 0.6 & 3.5 & & 0.7 & 5.5 & 5.1 & 7.2 & 1.3 & 4.4\\
\hspace{1em}3 & 0.0 & 4.2 & 0.0 & 6.4 & 0.5 & 4.0 & & \textbf{94.0} & \textbf{23.4} & \textbf{72.2} & \textbf{19.5} & \textbf{15.0} & \textbf{10.3}\\
\hspace{1em}4 & 1.3 & 6.9 & 2.9 & 21.2 & 3.7 & 6.5 & & 2.3 & 23.8 & 15.3 & 23.0 & 43.7 & 17.7\\
\hspace{1em}5 & 16.5 & 19.5 & 21.6 & 22.3 & 27.9 & 13.8 & & 2.3 & 19.4 & 5.9 & 19.3 & 35.1 & 16.8\\
\hspace{1em}6 & \textbf{78.8} & \textbf{20.7} & \textbf{68.0} & \textbf{17.6} & \textbf{66.9} & \textbf{23.0} & & 0.0 & 6.2 & 0.3 & 5.0 & 4.4 & 7.3\\
\hspace{1em} (None, N)& 3.4 & 59.0 & 7.5 & 74.7 & 0.0 & 54.1 & & 0.7 & 81.6 & 1.1 & 77.7 & 0.0 & 60.1\\
\addlinespace[0.3em]
\hline
\multicolumn{1}{c}{} & \multicolumn{6}{c}{Scenario 2} & \multicolumn{1}{c}{}  & \multicolumn{6}{c}{Scenario 7}   \\
\hspace{1em}1 & 0.0 & 3.2 & 0.0 & 3.6 & 0.3 & 3.4 & & 1.7 & 5.7 & 9.8 & 10.9 & 3.8 & 5.5\\
\hspace{1em}2 & 0.0 & 3.5 & 0.0 & 4.4 & 0.4 & 3.9 & & \textbf{77.9} & \textbf{20.6} & \textbf{71.7} & \textbf{20.4} & \textbf{20.7} & \textbf{12.2}\\
\hspace{1em}3 & 0.0 & 3.6 & 0.0 & 7.1 & 0.8 & 4.2 & & 16.6 & 23.1 & 16.2 & 22.5 & 36.0 & 16.6\\
\hspace{1em}4 & 3.1 & 15.4 & 6.7 & 21.4 & 6.7 & 8.0 & & 2.7 & 17.5 & 1.5 & 15.7 & 31.2 & 15.2\\
\hspace{1em}5 & \textbf{82.6} & \textbf{23.1} & \textbf{78.8} & \textbf{21.6} & \textbf{38.9} & \textbf{17.1} & & 0.1 & 6.0 & 0.8 & 5.3 & 5.1 & 7.6\\
\hspace{1em}6 & 9.4 & 19.8 & 5.6 & 16.2 & 52.9 & 20.1 & & 0.1 & 2.3 & 0.0 & 0.4 & 3.2 & 6.0\\
\hspace{1em} (None, N)& 4.9 & 68.5 & 8.9 & 74.2 & 0.0 & 56.7 & & 0.9 & 75.2 & 0.0 & 75.2 & 0.0 & 63.1\\
\addlinespace[0.3em]
\hline
\multicolumn{1}{c}{} & \multicolumn{6}{c}{Scenario 3} & \multicolumn{1}{c}{}  & \multicolumn{6}{c}{Scenario 8}   \\
\hspace{1em}1 & 0.0 & 3.4 & 0.0 & 4.1 & 0.0 & 3.5 & & 0.4 & 3.5 & 0.5 & 8.5 & 0.8 & 4.2\\
\hspace{1em}2 & 0.0 & 3.6 & 0.1 & 5.1 & 0.3 & 3.9 & & \textbf{48.9} & \textbf{22.4} & \textbf{41.9} & \textbf{18.4} & \textbf{23.2} & \textbf{12.8}\\
\hspace{1em}3 & 0.2 & 3.9 & 0.9 & 9.1 & 0.7 & 4.8 & & \textbf{49.9} & \textbf{23.7} & \textbf{53.9} & \textbf{20.9} & \textbf{28.0} & \textbf{13.9}\\
\hspace{1em}4 & \textbf{90.0} & \textbf{23.5} & \textbf{87.0} & \textbf{22.1} & \textbf{19.9} & \textbf{11.5} & & 0.6 & 20.2 & 2.3 & 18.2 & 34.3 & 16.1\\
\hspace{1em}5 & 3.9 & 23.1 & 2.1 & 20.2 & 45.7 & 17.9 & & 0.2 & 8.4 & 1.4 & 7.9 & 11.3 & 9.3\\
\hspace{1em}6 & 4.7 & 19.3 & 2.4 & 14.8 & 33.3 & 15.4 & & 0.0 & 2.9 & 0.0 & 1.0 & 2.4 & 4.8\\
\hspace{1em}(None, N)& 1.2 & 76.8 & 7.5 & 75.5 & 0.1 & 57.0 & & 0.0 & 81.1 & 0.0 & 74.8 & 0.0 & 61.1\\
\addlinespace[0.3em]
\hline
\multicolumn{1}{c}{} & \multicolumn{6}{c}{Scenario 4} & \multicolumn{1}{c}{}  & \multicolumn{6}{c}{Scenario 9}   \\
\hspace{1em}1 & 0.0 & 3.2 & 0.0 & 3.5 & 0.3 & 3.5 & & 0.0 & 3.3 & 0.1 & 3.5 & 2.2 & 5.6\\
\hspace{1em}2 & 0.0 & 3.4 & 0.1 & 4.9 & 0.3 & 3.7 & & 0.2 & 13.5 & 0.4 & 3.9 & 30.6 & 14.6\\
\hspace{1em}3 & 0.0 & 3.7 & 0.1 & 13.9 & 1.0 & 4.6 & & \textbf{83.2} & \textbf{23.2} & \textbf{5.3} & \textbf{4.3} & \textbf{47.3} & \textbf{19.4}\\
\hspace{1em}4 & 12.4 & 14.3 & 9.2 & 22.3 & 23.7 & 12.5 &  & 2.4 & 24.0 & 29.1 & 23.4 & 10.7 & 10.1\\
\hspace{1em}5 & \textbf{80.1} & \textbf{21.7} & \textbf{80.2} & \textbf{20.6} & \textbf{71.4} & \textbf{24.3} & & 4.9 & 23.9 & 27.4 & 22.9 & 5.4 & 8.3\\
\hspace{1em}6 & 3.6 & 7.7 & 2.8 & 7.0 & 3.2 & 7.9 &  & 9.3 & 23.0 & 37.0 & 22.4 & 3.5 & 7.7\\
\hspace{1em} (None, N)& 3.9 & 54.0 & 7.6 & 72.2 & 0.1 & 56.5 & & 0.0 & 110.9 & 0.7 & 80.4 & 0.3 & 65.7\\
\addlinespace[0.3em]
\hline
\multicolumn{1}{c}{} & \multicolumn{6}{c}{Scenario 5} & \multicolumn{1}{c}{}  & \multicolumn{6}{c}{Scenario 10}   \\
\hspace{1em}1 & 0.0 & 3.4 & 0.0 & 3.6 & 0.3 & 3.4 & & 0.0 & 9.2 & 0.0 & 13.0 & 0.7 & 10.6\\
\hspace{1em}2 & 0.0 & 3.8 & 0.0 & 5.6 & 0.7 & 3.8 & & 0.0 & 6.2 & 0.1 & 5.5 & 1.9 & 8.7\\
\hspace{1em}3 & 0.3 & 8.7 & 0.7 & 15.9 & 1.5 & 5.1 & & 0.0 & 3.5 & 0.1 & 2.1 & 10.5 & 10.9\\
\hspace{1em}4 & \textbf{94.3} & \textbf{23.3} & \textbf{92.1} & \textbf{22.8} & \textbf{53.6} & \textbf{20.3} & & 0.2 & 1.6 & 0.3 & 0.6 & 13.4 & 11.7\\
\hspace{1em}5 & 3.4 & 20.3 & 1.7 & 19.9 & 40.6 & 17.8 & & 0.0 & 0.9 & 0.1 & 0.1 & 7.7 & 10.0\\
\hspace{1em}6 & 0.1 & 7.5 & 0.2 & 5.9 & 3.3 & 7.0 & & 0.0 & 0.7 & 0.0 & 0.0 & 6.1 & 9.1\\
\hspace{1em}(None, N) & 1.9 & 67.0 & 5.3 & 73.6 & 0.0 & 57.4 & & \textbf{99.8} & \textbf{22.0} & \textbf{99.4} & \textbf{21.3} & \textbf{59.7} & \textbf{61.0}\\
\bottomrule
\end{tabular}
\end{center}
\endgroup{}
\end{table}

\setcounter{page}{1}
\renewcommand{\thepage}{S\arabic{page}}%
\setcounter{section}{0}
\renewcommand{\thesection}{S\arabic{section}}%
\setcounter{table}{0}
\renewcommand{\thetable}{S\arabic{table}}%
\setcounter{figure}{0}
\renewcommand{\thefigure}{S\arabic{figure}}%

\baselineskip=24pt
\begin{center}
{\Large \bf Supplementary Materials for ``DEMO: Dose Exploration, Monitoring, and Optimization Using
a Biological Mediator for Clinical Outcomes''}
\end{center}
\begin{center}
{\bf Cheng-Han Yang$^1$, Peter F. Thall$^2$, Ruitao Lin$^{2*}$}
\end{center}

\begin{center}

$^1$Department of Biostatistics, The University of Texas Health Science Center at Houston,\\
 Houston, TX 77030, USA\\

$^2$Department of Biostatistics, The University of Texas MD Anderson Cancer Center,\\
Houston, Texas 77030, U.S.A.\\

{$^*$Corresponding author: rlin@mdanderson.org}
\vspace{2mm}

\end{center}

\section{Posterior Model Probability for $Y_B$}

For patient $i$, $i=1,\ldots,n$, the probability model for an individual  outcome $Y_B(i)$ at dose $d(i) \in \{d_1,\ldots,d_J\}$ under  $\mathcal{M}_j$, $j=1,\ldots,J$ can be described as follows. 
\begin{eqnarray*}
	Y_{B}(i)\mid d(i) & \sim & \mathcal{N}(\mu_{B}(d(i)),\sigma_{B}^{2}),\\
	\mu_{B}(d(i)) & = & \mu_{-}I(d(i)<d_{j})+\mu_{+}I(d(i)\geq d_{j}),\\
	\mu_{-} & \sim & \mathcal{N}(m_{\mu_{-}},\nu_{\mu}^{2})\\
	\mu_{+} & \sim & \mathcal{N}(m_{\mu_{+}},\nu_{\mu}^{2})\\
	\zeta_{B}=\sigma_{B}^{-2} & \sim & {\rm Gamma}(a_{\sigma},b_{\sigma}).
\end{eqnarray*}
Without loss of generality, we take $\nu_{\mu}^{-2}$ as $n_0 \zeta_B$. By defining $\tau_B = d_j$, we partition the observations of the biological endpoint into inactive (-) and active (+) groups known as $L_j$ and $R_j$, respectively. Specifically, $L_j$ is the collection of all $Y_{B}(i)$ values for which $z(i)$ is less than $d_j$, while $R_j$ comprises all $y_{B}(i)$ values for which $d(i)$ is greater than or equal to $d_j$ under $\mathcal{M}_j$. 

We use the symbol $|\cdot|$ to denote the cardinality of a set. The sample means of the biological endpoints in $L_j$ and $R_j$ are represented by $\bar{y}_{L_j}$ and $\bar{y}_{R_j}$, respectively. Additionally, we denote the sample variances of the biological endpoints in  $L_j$ and $R_j$ as $s_{L_j}^2$ and $s_{R_j}^2$, respectively. Specifically, $s_{L_j}^2$ is calculated as $\sum_{i: d(i) < d_j} (y_{B}(i)-\bar{y}_{L_j})^2/ |L_j|$, and $s_{R_j}^2$ is calculated as $\sum_{i: d(i) \geq d_j} (y_{B}(i)-\bar{y}_{R_j})^2 / |R_j|$. Then, the posterior model probability $\text{Pr}(\mathcal{M}_j \mid \mathcal{D})$  can be calculated as  
\begin{equation}
 \text{Pr}(\mathcal{M}_j \mid \mathcal{D})  = \Pr(\mathcal{M}_j) \sqrt{\frac{1}{ |L_j|+n_0 }} \sqrt{\frac{1}{ |R_j|+n_0 }}  \frac{\Gamma(\tilde{a}_{\sigma})}{ {\tilde{b}_{\sigma}}^{\tilde{a}_{\sigma}}},
\end{equation}
where 
\begin{align*}
\tilde{a}_{\sigma} &= a_{\sigma} + \frac{N_1}{2} \\
\tilde{b}_{\sigma} &= b_{\sigma} + \frac{1}{2} |L_j| s_{L_j}^2 + \frac{1}{2} |R_j| s_{R_j}^2 + \frac{1}{2} \frac{|L_j| n_0}{|L_j| + n_0} (\bar{y}_{L_j}-m_{\mu_{-}})^2 + \frac{1}{2} \frac{|R_j| n_0}{|R_j| + n_0} (\bar{y}_{R_j}-m_{\mu_{+}})^2.
\end{align*}

The detailed derivation steps of the closed form of $\text{Pr}(\mathcal{M}_j \mid \mathcal{D})$ is given as follows. 
Let $\phi(\mu,\xi\mid Y)$ denote the likelihood of observing $Y$ from a normal distribution  mean $\mu$ and precision $\xi$.
Under model $\mathcal{M}_j$, the joint posterior distribution can be calculated by 
\begin{equation}
\label{joint_dist}
\begin{aligned}
\hspace*{-1.4cm}
    & p(\mu_{-}, \mu_{+}, \zeta_B \mid \mathcal{D}, \mathcal{M}_j)\\
     &= \prod_{ \{i: d(i)<d_j \}} \phi(\mu_{-}, \zeta_B| Y_B(i)) \pi(\mu_{-}) \prod_{\{i: d(i) \geq d_j \}} \phi(\mu_{+}, \zeta_B| Y_B(i) ) \pi(\mu_{+}) \pi(\zeta_B) \\
    = &\frac{\zeta_B^{\frac{|L_j|}{2}}}{(2\pi)^{\frac{|L_j|}{2}}} \exp \left\{ -\frac{1}{2} |L_j| s_{L_j}^2 \zeta_B - \frac{1}{2}|L_j| \zeta_B (\bar{y}_{L_j}-\mu_{-})^2 \right\}  \frac{(n_0\zeta_B)^{\frac{1}{2}}}{(2\pi)^{\frac{1}{2}}} \exp\left\{ -\frac{1}{2} n_0 \zeta_B (\mu_{-}-m_{\mu_{-}})^2  \right\} \times \\
    & \frac{\zeta_B^{\frac{|R_j|}{2}}}{(2\pi)^{\frac{|R_j|}{2}}} \exp \left\{ -\frac{1}{2} |R_j| s_{R_j}^2 \zeta_B - \frac{1}{2}|R_j| \zeta_B (\bar{y}_{R_j}-\mu_{+})^2 \right\}  \frac{(n_0\zeta_B)^{\frac{1}{2}}}{(2\pi)^{\frac{1}{2}}} \exp\left\{ -\frac{1}{2} n_0 \zeta_B (\mu_{+}-m_{\mu_{+}})^2  \right\} \times \\
    & \frac{b_{\sigma}^{a_{\sigma}}}{\Gamma(a_{\sigma})} \zeta_B^{a_{\sigma}-1} \exp(-b_{\sigma} \zeta_B) ,
\end{aligned}
\end{equation}
 By integrating the first term in the  joint posterior distribution (\ref{joint_dist}) with respect to $\mu_{-}$, we have 
\begin{align*}
 &   \int \exp\left\{ -\frac{1}{2} |L_j| \zeta_B (\bar{y}_{L_j}-\mu_{-})^2 - \frac{1}{2} n_0 \zeta_B (\mu_{-}-m_{\mu_{-}})^2 \right\} d \mu_{-} \\
        &= \int \exp \left\{ -\frac{\zeta_B}{2} (|L_j|+n_0) \left( \mu_{-} - \frac{|L_j| \bar{y}_{L_j} + n_0 m_{\mu_{-}}}{|L_j| + n_0} \right)^2 \right\} \times\\
         &\hspace*{1cm} \exp\left\{ -\frac{\zeta_B}{2} (|L_j| \bar{y}_{L_j}^2+n_0 m_{\mu_{-}}^2) + \frac{\zeta_B}{2} (|L_j|+n_0) \left( \frac{|L_j| \bar{y}_{L_j} }{|L_j|+n_0} \right)^2 \right\} d \mu_{-} \\
         &= \int \exp \left\{ -\frac{\zeta_B}{2} (|L_j|+n_0) \left( \mu_{-} - \frac{|L_j| \bar{y}_{L_j} + n_0 m_{\mu_{-}}}{|L_j| + n_0} \right)^2 \right\} d \mu_{-} \times \\
         &\hspace*{1cm}  \exp \left\{ -\frac{\zeta_B}{2} \frac{|L_j| n_0}{|L_j| + n_0} (\bar{y}_{L_j}-m_{\mu_{-}})^2 \right\} \\
         &= \left( 2 \pi \cdot \frac{1}{(|L_j|+n_0) \zeta_B} \right)^{\frac{1}{2}} \exp \left\{ -\frac{\zeta_B}{2} \frac{|L_j| n_0}{|L_j| + n_0} (\bar{y}_{L_j}-m_{\mu_{-}})^2 \right\} .
\end{align*}
 
Similarly,  by integrating the second term in joint posterior distribution (\ref{joint_dist}) with respect to $\mu_{+}$, we have 
\begin{align*}
& \int \exp\left\{ -\frac{1}{2} |R_j| \zeta_B (\bar{y}_{R_j}-\mu_{+})^2 - \frac{1}{2} n_0 \zeta_B (\mu_{+}-m_{\mu_{+}})^2 \right\} d \mu_{+}\\ 
&= \left( 2 \pi \cdot \frac{1}{(|R_j|+n_0) \zeta_B} \right)^{\frac{1}{2}} \exp \left\{ -\frac{\zeta_B}{2} \frac{|R_j| n_0}{|R_j| + n_0} (\bar{y}_{R_j}-m_{\mu_{+}})^2 \right\} .
\end{align*}
As a result, the integral of the joint posterior distribution (\ref{joint_dist}) with respect to $\mu_{-}$ and $\mu_{+}$ can be rewritten as 
\begin{equation}
\hspace*{-0.7cm}
\begin{aligned}
& p(\zeta_B \mid \mathcal{D},\mathcal{M}_j) \\
    &= \frac{1}{(2 \pi)^{\frac{N_1+2}{2}}} \frac{b_{\sigma}^{a_{\sigma}}}{\Gamma(a_{\sigma})}  \zeta_B^{\frac{N_1+2}{2}+a_{\sigma}-1} \exp \left( -\frac{1}{2} |L_j| s_{L_j}^2 \zeta_B - \frac{1}{2} |R_j| s_{R_j}^2 \zeta_B - b_{\sigma} \zeta_B \right) \times \\
    &\hspace*{0.1cm} \exp \left\{ -\frac{\zeta_B}{2} \frac{|L_j| n_0}{|L_j| + n_0} (\bar{y}_{L_j}-m_{\mu_{-}})^2 -\frac{\zeta_B}{2} \frac{|R_j| n_0}{|R_j| + n_0} (\bar{y}_{R_j}-m_{\mu_{+}})^2 \right\} \times \\
    &\left( 2 \pi \cdot \frac{1}{(|L_j|+n_0) \zeta_B} \right)^{\frac{1}{2}} \left( 2 \pi \cdot \frac{1}{(|R_j|+n_0) \zeta_B} \right)^{\frac{1}{2}} \\
&= \frac{1}{(2 \pi)^{\frac{N_1}{2}}} \frac{b_{\sigma}^{a_{\sigma}}}{\Gamma(a_{\sigma)} } \sqrt{\frac{n_0}{ |L_j|+n_0 }} \sqrt{\frac{n_0}{ |R_j|+n_0 }} \zeta_B^{\tilde{a}_{\sigma}-1} \exp \left\{- \tilde{b}_{\sigma} \zeta_B \right\}.
\end{aligned}
\end{equation}
\vspace{0.05in}

The probability $\text{Pr}(\mathcal{M}_j \mid \mathcal{D})$ can be obtained by Bayes theorem; that is,
\begin{align*}
\text{Pr}(\mathcal{M}_j \mid \mathcal{D}) &\propto \text{Pr}(\mathcal{M}_j) \int p(\zeta_B \mid \mathcal{D}, \mathcal{M}_j ) d \zeta_B \\
&= \text{Pr}(\mathcal{M}_j)\frac{1}{(2 \pi)^{\frac{N_1}{2}}} \frac{b_{\sigma}^{a_{\sigma}}}{\Gamma(a_{\sigma)} } \sqrt{\frac{n_0}{ |L_j|+n_0 }} \sqrt{\frac{n_0}{ |R_j|+n_0 }}   \int \zeta_B^{\tilde{a}_{\sigma}-1} \exp \left\{- \tilde{b}_{\sigma} \zeta_B \right\} d \zeta_B \\
&= \text{Pr}(\mathcal{M}_j)\frac{1}{(2 \pi)^{\frac{N_1}{2}}} \frac{b_{\sigma}^{a_{\sigma}}}{\Gamma(a_{\sigma)} }  \sqrt{\frac{n_0}{ |L_j|+n_0 }} \sqrt{\frac{n_0}{ |R_j|+n_0 }} \frac{\Gamma(\tilde{a}_{\sigma})}{ {\tilde{b}_{\sigma}}^{\tilde{a}_{\sigma}}}\\
&\propto \text{Pr}(\mathcal{M}_j)  \sqrt{\frac{1}{ |L_j|+n_0 }} \sqrt{\frac{1}{ |R_j|+n_0 }}   \frac{\Gamma(\tilde{a}_{\sigma})}{ {\tilde{b}_{\sigma}}^{\tilde{a}_{\sigma}}}.
\end{align*}

\newpage 

\section{Prior Specification}

\noindent
{\it Step function for $Y_B$}

For the first model of  $Y_B$, the hyperparameters are set as $m_{\mu_{-}}=0, m_{\mu_{+}}=0.5, a_{\sigma}=0.01, b_{\sigma}=0.01$, and $n_0=0.1$. Note that the hyperparameter $n_0$ is explicitly defined in Section S1 and can be treated as the prior sample size based on our prior knowledge. 

\vspace{0.15in}
\noindent
{\it Logistic regression model for $Y_T$}

To calibrate the prior parameters for the logistic regression model for $Y_T$ in stage 1, we ensure that the toxicity rate $\pi_T(d)$ lies between  0.001 and 0.999 with a prior probability greater than 0.90. As a result, the hyperparameters are set as $m_{\alpha_0}=-2, \nu_{\alpha_0}^2=10, m_{\alpha_1}=-0.693$ (i.e., the prior mean of the slope is $\exp(-0.693)=0.5$), and $\nu_{\alpha_1}^2=5$. 

\vspace{0.15in}
\noindent
{\it E$_{\max}$ model for $Y_B$}
We specify a non-informative prior for the the E$_{\max}$ model for $Y_B$, where we take $m_{\gamma}=0, \nu_{\gamma}^2=10, a_{\gamma_1}=1, b_{\gamma_1}=0.25, a_{\gamma_2}=0.1, b_{\gamma_2}=0.25, a_{\gamma_3}=0.75, b_{\gamma_3}=0.25, a_{\sigma}=0.1$, and $b_{\sigma}=0.1$. Under this prior, $\gamma_2$ ranges from 0 to 50, while $\gamma_3$ ranges from 0 to 10.


\vspace{0.15in}
\noindent
{\it Logistic regression models for $Y_T\mid Y_B$ and $Y_E \mid Y_B$}

Following the approach used in specifying hyperparameters for the marginal logistic regression model of $Y_T$, we select hyperparameter values to ensure that the prior probabilities of both $\pi_T(d)$ and $\pi_R(d)$ falling within the 0.001 to 0.999 range exceed 0.90. As a result,  the hyperparameters are specified as follows: $m_{\alpha_1}=-0.693$, $\nu_{\alpha_1}^2=5$, $m_{\alpha_2}=-2.302$, $\nu_{\alpha_2}^2=5$, $m_{\beta_1}=0$, $\nu_{\beta_1}^2=5$, $m_{\beta_2}=0$, $\nu_{\beta_2}^2=5$, $m_{\beta_3}=0$, $\nu_{\beta_3}^2=5$, $\mu_{\alpha_0}=-2$, $\mu_{\beta_0}=0$, $\nu_{\alpha_0}^2=10$, $\nu_{\beta_0}^2=5$, and $\rho_0=0.2$. 

\vspace{0.15in}
\noindent
{\it Weibull regression model for $Y_S\mid Y_T, Y_B, Y_R$}

We specify non-informative priors for the parameters associated in the Weibull regression model (4). In particular, we take $a_{\rho}=0.1$ and $b_{\rho}=0.1$, $\nu_{\lambda}^2=100$, and $\nu_{\eta}^2=100$.



\section*{Sensitivity Analysis}

We conducted a sensitivity analysis to evaluate the performance of DEMO under various hyperparameter configurations. Keeping other hyperparameters unchanged as in our main simulation study, we explored five additional settings, each involved changing the values of the hyperparameters for a single model at a time.

\begin{itemize}
\item Hyperparameter setting 1: In the  E$_{max}$ model  for $Y_B$, we took $m_{\gamma}=0, \nu_{\gamma}^2=10, a_{\gamma_1}=0.1875, b_{\gamma_1}=0.125, a_{\gamma_2}=0.075, b_{\gamma_2}=0.125, a_{\gamma_3}=0.3125, b_{\gamma_3}=0.125, a_{\sigma}=0.1$, and $b_{\sigma}=0.1$.
\item Hyperparameter setting 2: In the  stage 1 logistic regression model for $Y_T$,  we enlarged the prior variance and  took $m_{\alpha_0}=-2,  \nu_{\alpha_0}^2=50, m_{\alpha_1}=-0.693$, and $\nu_{\alpha_1}^2=50$.  
\item Hyperparameter setting 3: In  the logistic regression model for $Y_T\mid Y_B$, we enlarged the prior variance and   set $m_{\alpha_2}=-2.302$ and $\nu_{\alpha_2}^2=50$. 
\item Hyperparameter setting 4: In the logistic regression model for $Y_R \mid Y_B$, we enlarged the prior variance and set $\mu_{\beta_0}=0$, $\nu_{\beta_0}^2=50$, $m_{\beta_l}=0$ and $\nu_{\beta_l}^2=50$ for $l=1,2,3$.
\item Hyperparameter setting 5: In the Weibull regression model for $Y_S \mid Y_T, Y_B, Y_R$, we took $a_{\rho}=0.01$, $b_{\rho}=0.01$, $\nu_{\lambda}^2=1000$, and $\nu_{\eta}^2=1000$.
\end{itemize}

The results, presented in Table \ref{tab:sen1to5} and \ref{tab:sen6to10}, demonstrate that the operating characteristics of DEMO are generally robust against various settings of the hyperparameters. This suggests that provided the prior distributions are non-informative, characterized by a reasonably large variance, the DEMO design is capable of achieving robust performance.

\section{Impact of $(L,K,\kappa)$}


In this section, we examine the performance of DEMO  by investigating  multiple choices of $(L,K,\kappa)$. The purpose of $L$ is used to decide the minimum number of acceptable doses   that should be studied in stage 3 (given that there are more than $L$ acceptable doses at the end of stage 2). However, in some cases, there are multiple better performing doses when  efficacy plateaus at a lower dose. In such situations, $(K,\kappa)$ takes effect. Specifically,  the parameter $\kappa$ determines the strength of the evidence required, and at most $K$ dose levels are chosen into stage 3.   When $\kappa =0$, it means that at most $L$ doses will be selected, with no further consideration given to the remaining
$K-L$ doses. 


As shown in Table \ref{tab:sen_k1k2kappa1to5} and \ref{tab:sen_k1k2kappa6to10}, we considered different combinations of $L$, $K$, and $\kappa$. Generally speaking,  increasing $L$, $K$, and $\kappa$ results in a higher probability of correctly selecting the OTD, but this comes at the cost of enrolling a larger number of patients. However, the increase in the average total sample size due to a larger $(L,K,\kappa)$  is not substantial. When the patterns of $Y_R$ and $Y_S$ are more closely aligned, meaning the OBD is equal to or nearly equal to the OTD, employing smaller values of  $L$, $K$, and $\kappa$ can nearly achieve optimal performance while requiring a smaller sample size. This is evident in scenarios 1–5 and 7–8. On the other hand, when $Y_R$ lacks strong predictive power for $Y_S$ -- that is, when the OTD and OBD 
significantly differ, as demonstrated in scenario 9 -- opting for smaller values of  $L$, $K$, and $\kappa$  may result in inferior performance. 
From our extensive simulation studies, we found that  setting $(L,K,\kappa)=(2,3,0.3)$ for $J\leq 5$ doses and using $(L,K,\kappa)=(3,4,0.3)$ for  $J \geq 6$ doses 
typically achieves a good balance between a robustly high correct selection percentage and a reasonable sample size.

\begin{table}
\centering\begingroup\fontsize{8}{10}\selectfont
\caption{Selection percentages (Sel\%) and number of patients (Pts), under scenarios 1 to 5,  for the DEMO design using different hyperparameter settings. The optimal therapeutical dose under each scenario is given in boldface. ``None'' is the percentage of trials without selecting any dose, and ``N''  is the average total sample size.}
\label{tab:sen1to5}
\begin{tabular}[t]{crrrrrrrrrr}
\toprule
\multicolumn{1}{l}{Hyperparameter Setting} & \multicolumn{2}{c}{1} & \multicolumn{2}{c}{2} & \multicolumn{2}{c}{3} & \multicolumn{2}{c}{4} & \multicolumn{2}{c}{5} \\
\cmidrule(l{3pt}r{3pt}){2-3} \cmidrule(l{3pt}r{3pt}){4-5} \cmidrule(l{3pt}r{3pt}){6-7} \cmidrule(l{3pt}r{3pt}){8-9} \cmidrule(l{3pt}r{3pt}){10-11}
Dose & Sel\% & Pts & Sel\% & Pts & Sel\% & Pts & Sel\% & Pts & Sel\% & Pts \\
\midrule
\addlinespace[0.3em]
\multicolumn{11}{c}{\textbf{Scenario 1}}\\
\hspace{1em}1 & 0.0 & 3.8 & 0.0 & 3.8 & 0.0 & 3.8 & 0.0 & 3.8 & 0.0 & 3.8\\
\hspace{1em}2 & 0.0 & 4.0 & 0.0 & 4.0 & 0.0 & 4.0 & 0.0 & 4.0 & 0.0 & 4.0\\
\hspace{1em}3 & 0.0 & 4.2 & 0.0 & 4.3 & 0.0 & 4.2 & 0.0 & 4.2 & 0.0 & 4.2\\
\hspace{1em}4 & 1.3 & 7.0 & 1.3 & 7.4 & 1.7 & 6.9 & 0.5 & 6.5 & 1.3 & 6.9\\
\hspace{1em}5 & 15.9 & 19.4 & 20.0 & 19.5 & 19.7 & 19.5 & 16.0 & 19.1 & 16.5 & 19.5\\
\textbf{\hspace{1em}6} & \textbf{79.3} & \textbf{20.7} & \textbf{73.7} & \textbf{19.1} & \textbf{73.8} & \textbf{20.5} & \textbf{79.1} & \textbf{20.7} & \textbf{78.9} & \textbf{20.7}\\
\hspace{1em} (None, N) & 3.5 & 59.1 & 5.0 & 58.2 & 4.8 & 58.8 & 4.4 & 58.3 & 3.3 & 59.0\\
\addlinespace[0.3em]
\hline
\multicolumn{11}{c}{\textbf{Scenario 2}}\\
\hspace{1em}1 & 0.0 & 3.2 & 0.0 & 3.2 & 0.0 & 3.2 & 0.0 & 3.2 & 0.0 & \vphantom{1} 3.2\\
\hspace{1em}2 & 0.0 & 3.5 & 0.0 & 3.5 & 0.0 & 3.5 & 0.0 & 3.5 & 0.0 & 3.5\\
\hspace{1em}3 & 0.0 & 3.6 & 0.0 & 3.6 & 0.0 & 3.6 & 0.0 & 3.5 & 0.0 & 3.6\\
\hspace{1em}4 & 4.2 & 15.6 & 4.4 & 15.6 & 3.0 & 15.4 & 2.2 & 14.2 & 3.0 & 15.4\\
\textbf{\hspace{1em}5} & \textbf{81.1} & \textbf{23.1} & \textbf{80.6} & \textbf{22.9} & \textbf{84.0} & \textbf{23.1} & \textbf{77.4} & \textbf{22.8} & \textbf{82.9} & \textbf{23.1}\\
\hspace{1em}6 & 9.6 & 19.8 & 9.2 & 18.4 & 8.5 & 19.6 & 15.6 & 19.8 & 9.3 & 19.8\\
\hspace{1em} (None, N) & 5.1 & 68.6 & 5.8 & 67.2 & 4.5 & 68.4 & 4.8 & 67.0 & 4.8 & 68.5\\
\addlinespace[0.3em]
\hline
\multicolumn{11}{c}{\textbf{Scenario 3}}\\
\hspace{1em}1 & 0.0 & 3.4 & 0.0 & 3.4 & 0.0 & 3.4 & 0.0 & 3.4 & 0.0 & \vphantom{1} 3.4\\
\hspace{1em}2 & 0.0 & 3.6 & 0.0 & 3.6 & 0.0 & 3.6 & 0.0 & 3.6 & 0.0 & 3.6\\
\hspace{1em}3 & 0.2 & 3.9 & 0.2 & 3.9 & 0.2 & 3.9 & 0.1 & 3.9 & 0.2 & 3.9\\
\textbf{\hspace{1em}4} & \textbf{90.4} & \textbf{23.4} & \textbf{89.4} & \textbf{23.3} & \textbf{89.5} & \textbf{23.4} & \textbf{83.7} & \textbf{23.1} & \textbf{90.0} & \textbf{23.5}\\
\hspace{1em}5 & 3.5 & 23.1 & 4.2 & 22.8 & 4.1 & 23.1 & 6.0 & 23.1 & 3.7 & 23.1\\
\hspace{1em}6 & 4.7 & 19.3 & 4.7 & 18.5 & 5.1 & 19.1 & 9.0 & 19.3 & 4.8 & 19.3\\
\hspace{1em} (None, N) & 1.2 & 76.8 & 1.5 & 75.6 & 1.1 & 76.6 & 1.2 & 76.4 & 1.3 & 76.8\\
\addlinespace[0.3em]
\hline
\multicolumn{11}{c}{\textbf{Scenario 4}}\\
\hspace{1em}1 & 0.0 & 3.2 & 0.0 & 3.2 & 0.0 & 3.2 & 0.0 & 3.2 & 0.0 & 3.2\\
\hspace{1em}2 & 0.0 & 3.4 & 0.0 & 3.4 & 0.0 & 3.4 & 0.0 & 3.4 & 0.0 & 3.4\\
\hspace{1em}3 & 0.0 & 3.7 & 0.0 & 3.7 & 0.0 & 3.7 & 0.0 & 3.6 & 0.0 & 3.7\\
\hspace{1em}4 & 13.2 & 14.4 & 13.9 & 14.4 & 12.8 & 14.3 & 10.9 & 13.9 & 12.4 & 14.3\\
\textbf{\hspace{1em}5} & \textbf{77.4} & \textbf{21.6} & \textbf{78.6} & \textbf{21.2} & \textbf{81.0} & \textbf{21.7} & \textbf{79.9} & \textbf{21.6} & \textbf{80.3} & \textbf{21.7}\\
\hspace{1em}6 & 5.4 & 7.8 & 2.9 & 7.0 & 2.5 & 7.6 & 3.6 & 7.7 & 3.6 & 7.7\\
\hspace{1em} (None, N) & 4.0 & 54.1 & 4.6 & 52.8 & 3.7 & 53.8 & 5.6 & 53.5 & 3.7 & 54.0\\
\addlinespace[0.3em]
\hline
\multicolumn{11}{c}{\textbf{Scenario 5}}\\
\hspace{1em}1 & 0.0 & 3.4 & 0.0 & 3.4 & 0.0 & 3.4 & 0.0 & 3.4 & 0.0 & 3.4\\
\hspace{1em}2 & 0.0 & 3.8 & 0.0 & 3.9 & 0.0 & 3.8 & 0.0 & 3.8 & 0.0 & 3.8\\
\hspace{1em}3 & 0.5 & 8.9 & 0.2 & 8.7 & 0.3 & 8.7 & 0.3 & 8.2 & 0.3 & 8.7\\
\textbf{\hspace{1em}4} & \textbf{93.5} & \textbf{23.3} & \textbf{93.4} & \textbf{23.2} & \textbf{94.7} & \textbf{23.4} & \textbf{90.8} & \textbf{23.0} & \textbf{94.3} & \textbf{23.3}\\
\hspace{1em}5 & 3.8 & 20.3 & 3.5 & 19.3 & 3.2 & 20.3 & 6.1 & 20.3 & 3.4 & 20.3\\
\hspace{1em}6 & 0.1 & 7.5 & 0.1 & 7.0 & 0.1 & 7.4 & 0.1 & 7.4 & 0.1 & 7.5\\
\hspace{1em} (None, N) & 2.1 & 67.2 & 2.8 & 65.6 & 1.7 & 67.0 & 2.7 & 66.1 & 1.9 & 67.0\\
\bottomrule
\end{tabular}
\endgroup{}
\end{table}

\begin{table}
\centering\begingroup\fontsize{8}{10}\selectfont
\caption{Selection percentages (Sel\%) and number of patients (Pts), under scenarios 1 to 5,  for the DEMO design using different hyperparameter settings. The optimal therapeutical dose under each scenario is given in boldface. ``None'' is the percentage of trials without selecting any dose, and ``N''  is the average total sample size.}
\label{tab:sen6to10}
\begin{tabular}[t]{crrrrrrrrrr}
\toprule
\multicolumn{1}{l}{Hyperparameter Setting} & \multicolumn{2}{c}{1} & \multicolumn{2}{c}{2} & \multicolumn{2}{c}{3} & \multicolumn{2}{c}{4} & \multicolumn{2}{c}{5} \\
\cmidrule(l{3pt}r{3pt}){2-3} \cmidrule(l{3pt}r{3pt}){4-5} \cmidrule(l{3pt}r{3pt}){6-7} \cmidrule(l{3pt}r{3pt}){8-9} \cmidrule(l{3pt}r{3pt}){10-11}
\cmidrule(l{3pt}r{3pt}){2-3} \cmidrule(l{3pt}r{3pt}){4-5} \cmidrule(l{3pt}r{3pt}){6-7} \cmidrule(l{3pt}r{3pt}){8-9} \cmidrule(l{3pt}r{3pt}){10-11}
Dose & Sel\% & Pts & Sel\% & Pts & Sel\% & Pts & Sel\% & Pts & Sel\% & Pts \\
\midrule
\addlinespace[0.3em]
\multicolumn{11}{c}{\textbf{Scenario 6}}\\
\hspace{1em}1 & 0.0 & 3.2 & 0.0 & 3.2 & 0.0 & 3.2 & 0.0 & 3.2 & 0.0 & 3.2\\
\hspace{1em}2 & 1.2 & 5.6 & 0.6 & 5.5 & 0.7 & 5.5 & 0.6 & 5.5 & 0.7 & 5.5\\
\textbf{\hspace{1em}3} & \textbf{93.3} & \textbf{23.3} & \textbf{94.3} & \textbf{23.4} & \textbf{94.0} & \textbf{23.4} & \textbf{87.8} & \textbf{22.7} & \textbf{94.0} & \textbf{23.4}\\
\hspace{1em}4 & 2.2 & 23.8 & 2.0 & 23.6 & 2.4 & 23.8 & 5.4 & 23.7 & 2.6 & 23.8\\
\hspace{1em}5 & 2.3 & 19.4 & 2.3 & 18.5 & 2.2 & 19.4 & 4.9 & 19.4 & 2.0 & 19.4\\
\hspace{1em}6 & 0.0 & 6.3 & 0.0 & 5.9 & 0.0 & 6.2 & 0.0 & 6.2 & 0.0 & 6.2\\
\hspace{1em} (None, N) & 1.0 & 81.6 & 0.8 & 80.1 & 0.7 & 81.5 & 1.3 & 80.8 & 0.7 & 81.6\\
\addlinespace[0.3em]
\hline
\multicolumn{11}{c}{\textbf{Scenario 7}}\\
\hspace{1em}1 & 1.8 & 5.7 & 1.3 & 5.6 & 1.7 & 5.7 & 2.0 & 5.7 & 1.7 & 5.7\\
\textbf{\hspace{1em}2} & \textbf{77.8} & \textbf{20.6} & \textbf{76.1} & \textbf{20.3} & \textbf{77.9} & \textbf{20.6} & \textbf{77.3} & \textbf{20.6} & \textbf{77.9} & \textbf{20.6}\\
\hspace{1em}3 & 16.5 & 23.1 & 16.8 & 22.7 & 16.5 & 23.1 & 16.4 & 23.0 & 16.4 & 23.1\\
\hspace{1em}4 & 2.7 & 17.5 & 3.5 & 17.1 & 2.7 & 17.5 & 2.8 & 17.5 & 2.8 & 17.5\\
\hspace{1em}5 & 0.1 & 6.0 & 0.6 & 6.5 & 0.1 & 6.0 & 0.1 & 6.0 & 0.1 & 6.0\\
\hspace{1em}6 & 0.1 & 2.3 & 0.0 & 2.8 & 0.1 & 2.3 & 0.1 & 2.2 & 0.1 & 2.3\\
\hspace{1em} (None, N) & 1.0 & 75.2 & 1.7 & 75.0 & 1.0 & 75.2 & 1.3 & 75.1 & 1.0 & 75.2\\
\addlinespace[0.3em]
\hline
\multicolumn{11}{c}{\textbf{Scenario 8}}\\
\hspace{1em}1 & 0.4 & 3.5 & 0.3 & 3.5 & 0.4 & 3.5 & 0.4 & 3.5 & 0.4 & 3.5\\
\textbf{\hspace{1em}2} & \textbf{50.3} & \textbf{22.5} & \textbf{49.0} & \textbf{22.3} & \textbf{47.8} & \textbf{22.4} & \textbf{50.1} & \textbf{22.4} & \textbf{48.9} & \textbf{22.4}\\
\textbf{\hspace{1em}3} & \textbf{48.6} & \textbf{23.7} & \textbf{49.5} & \textbf{23.6} & \textbf{50.9} & \textbf{23.7} & \textbf{48.5} & \textbf{23.7} & \textbf{49.9} & \textbf{23.7}\\
\hspace{1em}4 & 0.5 & 20.2 & 0.5 & 19.8 & 0.7 & 20.2 & 0.8 & 20.2 & 0.6 & 20.2\\
\hspace{1em}5 & 0.2 & 8.4 & 0.1 & 8.7 & 0.2 & 8.4 & 0.2 & 8.4 & 0.2 & 8.4\\
\hspace{1em}6 & 0.0 & 2.9 & 0.0 & 3.4 & 0.0 & 3.0 & 0.0 & 2.9 & 0.0 & 2.9\\
\hspace{1em} (None, N) & 0.0 & 81.2 & 0.6 & 81.3 & 0.0 & 81.2 & 0.0 & 81.1 & 0.0 & 81.1\\
\addlinespace[0.3em]
\hline
\multicolumn{11}{c}{\textbf{Scenario 9}}\\
\hspace{1em}1 & 0.0 & 3.3 & 0.0 & 3.3 & 0.0 & 3.3 & 0.0 & 3.3 & 0.0 & 3.3\\
\hspace{1em}2 & 0.2 & 13.6 & 0.2 & 13.6 & 0.1 & 13.5 & 0.1 & 13.6 & 0.2 & 13.5\\
\textbf{\hspace{1em}3} & \textbf{82.6} & \textbf{23.2} & \textbf{83.4} & \textbf{23.2} & \textbf{82.7} & \textbf{23.2} & \textbf{78.9} & \textbf{22.8} & \textbf{83.0} & \textbf{23.2}\\
\hspace{1em}4 & 2.6 & 24.0 & 2.3 & 23.9 & 2.5 & 24.0 & 3.5 & 24.0 & 2.4 & 24.0\\
\hspace{1em}5 & 5.2 & 23.9 & 4.8 & 23.9 & 5.1 & 23.9 & 6.5 & 23.9 & 5.0 & 23.9\\
\hspace{1em}6 & 9.4 & 22.9 & 9.2 & 22.9 & 9.6 & 23.0 & 11.0 & 22.9 & 9.4 & 23.0\\
\hspace{1em} (None, N) & 0.0 & 110.9 & 0.1 & 110.8 & 0.0 & 110.9 & 0.0 & 110.4 & 0.0 & 110.9\\
\addlinespace[0.3em]
\hline
\multicolumn{11}{c}{\textbf{Scenario 10}}\\
\hspace{1em}1 & 0.0 & 9.2 & 0.1 & 9.7 & 0.0 & 9.2 & 0.0 & 9.2 & 0.0 & 9.2\\
\hspace{1em}2 & 0.0 & 6.2 & 0.0 & 5.1 & 0.0 & 6.2 & 0.0 & 6.2 & 0.0 & 6.2\\
\hspace{1em}3 & 0.0 & 3.5 & 0.0 & 3.3 & 0.0 & 3.5 & 0.0 & 3.5 & 0.0 & 3.5\\
\hspace{1em}4 & 0.0 & 1.6 & 0.1 & 1.6 & 0.2 & 1.6 & 0.1 & 1.6 & 0.2 & 1.6\\
\hspace{1em}5 & 0.0 & 0.9 & 0.2 & 1.2 & 0.0 & 0.9 & 0.0 & 0.9 & 0.0 & 0.9\\
\hspace{1em}6 & 0.0 & 0.7 & 0.0 & 1.1 & 0.0 & 0.7 & 0.1 & 0.7 & 0.0 & 0.7\\
Not found & \textbf{100.0} & \textbf{22.0} & \textbf{99.6} & \textbf{22.0} & \textbf{99.8} & \textbf{22.0} & \textbf{99.8} & \textbf{22.0} & \textbf{99.8} & \textbf{22.0}\\
\bottomrule
\end{tabular}
\endgroup{}
\end{table}

\begin{table}
\centering\begingroup\fontsize{7}{9}\selectfont
\caption{Selection percentages (Sel\%) and number of patients (Pts) at each dose for different settings of $(L,K,\kappa)$ under scenarios 1 to 5. The optimal therapeutical dose under each scenario is given in boldface. ``None'' is the percentage of trials without selecting any dose, and ``N''  is the average total sample size.}
\label{tab:sen_k1k2kappa1to5}
\begin{tabular}[t]{lrrrrrrrrrrrrrr}
\toprule
\multicolumn{1}{c}{} & \multicolumn{14}{c}{($L,K,\kappa$)} \\
\cmidrule(l{3pt}r{3pt}){2-15}
\multicolumn{1}{c}{} & \multicolumn{2}{c}{(2,3,0.3)} & \multicolumn{2}{c}{(2,4,0.3)} & \multicolumn{2}{c}{(2,6,0.3)} & \multicolumn{2}{c}{(3,4,0.0)} & \multicolumn{2}{c}{(3,4,0.2)} & \multicolumn{2}{c}{(3,4,0.3)} & \multicolumn{2}{c}{(3,6,0.3)} \\
\cmidrule(l{3pt}r{3pt}){2-3} \cmidrule(l{3pt}r{3pt}){4-5} \cmidrule(l{3pt}r{3pt}){6-7} \cmidrule(l{3pt}r{3pt}){8-9} \cmidrule(l{3pt}r{3pt}){10-11} \cmidrule(l{3pt}r{3pt}){12-13} \cmidrule(l{3pt}r{3pt}){14-15}
Dose & Sel\% & Pts & Sel\% & Pts & Sel\% & Pts & Sel\% & Pts & Sel\% & Pts & Sel\% & Pts & Sel\% & Pts\\
\midrule
\addlinespace[0.3em]
\multicolumn{15}{c}{\textbf{Scenario 1}}\\
\hspace{1em}1 & 0.0 & 3.8 & 0.0 & 3.8 & 0.0 & 3.8 & 0.0 & 3.8 & 0.0 & 3.8 & 0.0 & 3.8 & 0.0 & 3.8\\
\hspace{1em}2 & 0.0 & 4.0 & 0.0 & 4.0 & 0.0 & 4.0 & 0.0 & 4.0 & 0.0 & 4.0 & 0.0 & 4.0 & 0.0 & 4.0\\
\hspace{1em}3 & 0.0 & 4.2 & 0.0 & 4.2 & 0.0 & 4.2 & 0.0 & 4.2 & 0.0 & 4.2 & 0.0 & 4.2 & 0.0 & 4.2\\
\hspace{1em}4 & 1.3 & 6.6 & 1.3 & 6.6 & 1.3 & 6.6 & 1.3 & 6.9 & 1.3 & 6.9 & 1.3 & 6.9 & 1.3 & 6.9\\
\hspace{1em}5 & 16.4 & 19.5 & 16.4 & 19.5 & 16.4 & 19.5 & 16.5 & 19.5 & 16.5 & 19.5 & 16.5 & 19.5 & 16.5 & 19.5\\
\textbf{\hspace{1em}6} & \textbf{79.0} & \textbf{20.7} & \textbf{79.0} & \textbf{20.7} & \textbf{79.0} & \textbf{20.7} & \textbf{78.8} & \textbf{20.7} & \textbf{78.8} & \textbf{20.7} & \textbf{78.8} & \textbf{20.7} & \textbf{78.8} & \textbf{20.7}\\
\hspace{1em} (None, N) & 3.3 & 58.7 & 3.3 & 58.7 & 3.3 & 58.7 & 3.4 & 59.0 & 3.4 & 59.0 & 3.4 & 59.0 & 3.4 & 59.0\\
\hline
\addlinespace[0.3em]
\multicolumn{15}{c}{\textbf{Scenario 2}}\\
\hspace{1em}1 & 0.0 & 3.2 & 0.0 & 3.2 & 0.0 & 3.2 & 0.0 & 3.2 & 0.0 & 3.2 & 0.0 & 3.2 & 0.0 & \vphantom{1} 3.2\\
\hspace{1em}2 & 0.0 & 3.5 & 0.0 & 3.5 & 0.0 & 3.5 & 0.0 & 3.5 & 0.0 & 3.5 & 0.0 & 3.5 & 0.0 & 3.5\\
\hspace{1em}3 & 0.0 & 3.5 & 0.0 & 3.5 & 0.0 & 3.5 & 0.0 & 3.6 & 0.0 & 3.6 & 0.0 & 3.6 & 0.0 & 3.6\\
\hspace{1em}4 & 2.8 & 14.2 & 2.8 & 14.2 & 2.8 & 14.2 & 3.1 & 15.4 & 3.1 & 15.4 & 3.1 & 15.4 & 3.1 & 15.4\\
\textbf{\hspace{1em}5} & \textbf{83.5} & \textbf{23.1} & \textbf{83.5} & \textbf{23.1} & \textbf{83.5} & \textbf{23.1} & \textbf{82.6} & \textbf{23.1} & \textbf{82.6} & \textbf{23.1} & \textbf{82.6} & \textbf{23.1} & \textbf{82.6} & \textbf{23.1}\\
\hspace{1em}6 & 8.9 & 19.8 & 8.9 & 19.8 & 8.9 & 19.8 & 9.4 & 19.8 & 9.4 & 19.8 & 9.4 & 19.8 & 9.4 & 19.8\\
\hspace{1em} (None, N) & 4.8 & 67.3 & 4.8 & 67.3 & 4.8 & 67.3 & 4.9 & 68.5 & 4.9 & 68.5 & 4.9 & 68.5 & 4.9 & 68.5\\
\addlinespace[0.3em]
\hline
\multicolumn{15}{c}{\textbf{Scenario 3}}\\
\hspace{1em}1 & 0.0 & 3.4 & 0.0 & 3.4 & 0.0 & 3.4 & 0.0 & 3.4 & 0.0 & 3.4 & 0.0 & 3.4 & 0.0 & \vphantom{1} 3.4\\
\hspace{1em}2 & 0.0 & 3.6 & 0.0 & 3.6 & 0.0 & 3.6 & 0.0 & 3.6 & 0.0 & 3.6 & 0.0 & 3.6 & 0.0 & 3.6\\
\hspace{1em}3 & 0.2 & 3.9 & 0.2 & 3.9 & 0.2 & 3.9 & 0.2 & 3.9 & 0.2 & 3.9 & 0.2 & 3.9 & 0.2 & 3.9\\
\textbf{\hspace{1em}4} & \textbf{86.8} & \textbf{23.0} & \textbf{86.8} & \textbf{23.0} & \textbf{86.8} & \textbf{23.0} & \textbf{90.0} & \textbf{23.5} & \textbf{90.0} & \textbf{23.5} & \textbf{90.0} & \textbf{23.5} & \textbf{90.0} & \textbf{23.5}\\
\hspace{1em}5 & 5.3 & 23.1 & 5.3 & 23.1 & 5.3 & 23.1 & 3.9 & 23.1 & 3.9 & 23.1 & 3.9 & 23.1 & 3.9 & 23.1\\
\hspace{1em}6 & 6.7 & 19.2 & 6.7 & 19.2 & 6.7 & 19.2 & 4.7 & 19.3 & 4.7 & 19.3 & 4.7 & 19.3 & 4.7 & 19.3\\
\hspace{1em} (None, N) & 1.0 & 76.3 & 1.0 & 76.3 & 1.0 & 76.3 & 1.2 & 76.8 & 1.2 & 76.8 & 1.2 & 76.8 & 1.2 & 76.8\\
\addlinespace[0.3em]
\hline
\multicolumn{15}{c}{\textbf{Scenario 4}}\\
\hspace{1em}1 & 0.0 & 3.2 & 0.0 & 3.2 & 0.0 & 3.2 & 0.0 & 3.2 & 0.0 & 3.2 & 0.0 & 3.2 & 0.0 & 3.2\\
\hspace{1em}2 & 0.0 & 3.4 & 0.0 & 3.4 & 0.0 & 3.4 & 0.0 & 3.4 & 0.0 & 3.4 & 0.0 & 3.4 & 0.0 & 3.4\\
\hspace{1em}3 & 0.0 & 3.6 & 0.0 & 3.6 & 0.0 & 3.6 & 0.0 & 3.7 & 0.0 & 3.7 & 0.0 & 3.7 & 0.0 & 3.7\\
\hspace{1em}4 & 12.5 & 14.3 & 12.5 & 14.3 & 12.5 & 14.3 & 12.4 & 14.3 & 12.4 & 14.3 & 12.4 & 14.3 & 12.4 & 14.3\\
\textbf{\hspace{1em}5} & \textbf{80.1} & \textbf{21.7} & \textbf{80.1} & \textbf{21.7} & \textbf{80.1} & \textbf{21.7} & \textbf{80.1} & \textbf{21.7} & \textbf{80.1} & \textbf{21.7} & \textbf{80.1} & \textbf{21.7} & \textbf{80.1} & \textbf{21.7}\\
\hspace{1em}6 & 3.6 & 7.7 & 3.6 & 7.7 & 3.6 & 7.7 & 3.6 & 7.7 & 3.6 & 7.7 & 3.6 & 7.7 & 3.6 & 7.7\\
\hspace{1em} (None, N) & 3.8 & 53.9 & 3.8 & 53.9 & 3.8 & 53.9 & 3.9 & 54.0 & 3.9 & 54.0 & 3.9 & 54.0 & 3.9 & 54.0\\
\addlinespace[0.3em]
\hline
\multicolumn{15}{c}{\textbf{Scenario 5}}\\
\hspace{1em}1 & 0.0 & 3.4 & 0.0 & 3.4 & 0.0 & 3.4 & 0.0 & 3.4 & 0.0 & 3.4 & 0.0 & 3.4 & 0.0 & 3.4\\
\hspace{1em}2 & 0.0 & 3.8 & 0.0 & 3.8 & 0.0 & 3.8 & 0.0 & 3.8 & 0.0 & 3.8 & 0.0 & 3.8 & 0.0 & 3.8\\
\hspace{1em}3 & 0.3 & 8.2 & 0.3 & 8.3 & 0.3 & 8.3 & 0.3 & 8.6 & 0.3 & 8.7 & 0.3 & 8.7 & 0.3 & 8.7\\
\textbf{\hspace{1em}4} & \textbf{92.9} & \textbf{23.1} & \textbf{92.9} & \textbf{23.1} & \textbf{92.9} & \textbf{23.1} & \textbf{94.3} & \textbf{23.3} & \textbf{94.3} & \textbf{23.3} & \textbf{94.3} & \textbf{23.3} & \textbf{94.3} & \textbf{23.3}\\
\hspace{1em}5 & 4.3 & 20.3 & 4.3 & 20.3 & 4.3 & 20.3 & 3.4 & 20.3 & 3.4 & 20.3 & 3.4 & 20.3 & 3.4 & 20.3\\
\hspace{1em}6 & 0.3 & 7.4 & 0.3 & 7.4 & 0.3 & 7.4 & 0.1 & 7.5 & 0.1 & 7.5 & 0.1 & 7.5 & 0.1 & 7.5\\
\hspace{1em} (None, N) & 2.2 & 66.2 & 2.2 & 66.3 & 2.2 & 66.3 & 1.9 & 67.0 & 1.9 & 67.0 & 1.9 & 67.0 & 1.9 & 67.0\\
\bottomrule
\end{tabular}
\endgroup{}
\end{table}

\begin{table}
\centering\begingroup\fontsize{7}{9}\selectfont
\caption{Selection percentages (Sel\%) and number of patients (Pts) at each dose for different settings of $(L,K,\kappa)$ under scenarios 6 to 10. The optimal therapeutical dose under each scenario is given in boldface. ``None'' is the percentage of trials without selecting any dose, and ``N''  is the average total sample size.}
\label{tab:sen_k1k2kappa6to10}
\begin{tabular}[t]{lrrrrrrrrrrrrrr}
\toprule
\multicolumn{1}{c}{} & \multicolumn{14}{c}{($L,K,\kappa$)} \\
\cmidrule(l{3pt}r{3pt}){2-15}
\multicolumn{1}{c}{} & \multicolumn{2}{c}{(2,3,0.3)} & \multicolumn{2}{c}{(2,4,0.3)} & \multicolumn{2}{c}{(2,6,0.3)} & \multicolumn{2}{c}{(3,4,0.0)} & \multicolumn{2}{c}{(3,4,0.2)} & \multicolumn{2}{c}{(3,4,0.3)} & \multicolumn{2}{c}{(3,6,0.3)} \\
\cmidrule(l{3pt}r{3pt}){2-3} \cmidrule(l{3pt}r{3pt}){4-5} \cmidrule(l{3pt}r{3pt}){6-7} \cmidrule(l{3pt}r{3pt}){8-9} \cmidrule(l{3pt}r{3pt}){10-11} \cmidrule(l{3pt}r{3pt}){12-13} \cmidrule(l{3pt}r{3pt}){14-15}
Dose & Sel\% & Pts & Sel\% & Pts & Sel\% & Pts & Sel\% & Pts & Sel\% & Pts & Sel\% & Pts & Sel\% & Pts\\
\midrule
\addlinespace[0.3em]
\multicolumn{15}{c}{\textbf{Scenario 6}}\\
\hspace{1em}1 & 0.0 & 3.2 & 0.0 & 3.2 & 0.0 & 3.2 & 0.0 & 3.2 & 0.0 & 3.2 & 0.0 & 3.2 & 0.0 & 3.2\\
\hspace{1em}2 & 0.2 & 5.3 & 0.7 & 5.5 & 0.6 & 5.5 & 0.2 & 5.3 & 0.7 & 5.5 & 0.7 & 5.5 & 0.6 & 5.5\\
\textbf{\hspace{1em}3} & \textbf{80.5} & \textbf{21.7} & \textbf{85.0} & \textbf{22.3} & \textbf{85.1} & \textbf{22.3} & \textbf{89.5} & \textbf{22.8} & \textbf{93.1} & \textbf{23.3} & \textbf{94.0} & \textbf{23.4} & \textbf{94.1} & \textbf{23.4}\\
\hspace{1em}4 & 11.0 & 23.8 & 7.3 & 23.8 & 7.3 & 23.8 & 6.0 & 23.8 & 2.9 & 23.8 & 2.3 & 23.8 & 2.3 & 23.8\\
\hspace{1em}5 & 7.0 & 19.4 & 6.1 & 19.4 & 6.1 & 19.4 & 3.2 & 19.4 & 2.6 & 19.4 & 2.3 & 19.4 & 2.3 & 19.4\\
\hspace{1em}6 & 0.4 & 6.1 & 0.0 & 6.2 & 0.0 & 6.2 & 0.4 & 6.1 & 0.0 & 6.2 & 0.0 & 6.2 & 0.0 & 6.2\\
\hspace{1em} (None, N) & 0.9 & 79.6 & 0.9 & 80.4 & 0.9 & 80.4 & 0.7 & 80.7 & 0.7 & 81.5 & 0.7 & 81.6 & 0.7 & 81.6\\
\addlinespace[0.3em]
\hline
\multicolumn{15}{c}{\textbf{Scenario 7}}\\
\hspace{1em}1 & 0.6 & 5.0 & 1.7 & 5.7 & 2.1 & 5.9 & 0.6 & 5.0 & 1.7 & 5.7 & 1.7 & 5.7 & 2.1 & 5.9\\
\textbf{\hspace{1em}2} & \textbf{73.0} & \textbf{19.9} & \textbf{76.4} & \textbf{20.4} & \textbf{77.3} & \textbf{20.6} & \textbf{74.5} & \textbf{20.1} & \textbf{77.8} & \textbf{20.6} & \textbf{77.9} & \textbf{20.6} & \textbf{78.8} & \textbf{20.8}\\
\hspace{1em}3 & 21.0 & 23.0 & 18.0 & 23.1 & 16.8 & 23.1 & 19.6 & 23.0 & 16.7 & 23.1 & 16.6 & 23.1 & 15.4 & 23.1\\
\hspace{1em}4 & 3.8 & 17.4 & 2.8 & 17.5 & 2.7 & 17.5 & 3.7 & 17.4 & 2.7 & 17.5 & 2.7 & 17.5 & 2.6 & 17.5\\
\hspace{1em}5 & 0.6 & 5.9 & 0.1 & 6.0 & 0.1 & 6.1 & 0.6 & 5.9 & 0.1 & 6.0 & 0.1 & 6.0 & 0.1 & 6.1\\
\hspace{1em}6 & 0.1 & 2.3 & 0.1 & 2.3 & 0.1 & 2.3 & 0.1 & 2.3 & 0.1 & 2.3 & 0.1 & 2.3 & 0.1 & 2.3\\
\hspace{1em} (None, N) & 0.9 & 73.5 & 0.9 & 75.0 & 0.9 & 75.4 & 0.9 & 73.7 & 0.9 & 75.2 & 0.9 & 75.2 & 0.9 & 75.6\\
\addlinespace[0.3em]
\hline
\multicolumn{15}{c}{\textbf{Scenario 8}}\\
\hspace{1em}1 & 0.4 & 3.4 & 0.4 & 3.5 & 0.4 & 3.5 & 0.4 & 3.4 & 0.4 & 3.5 & 0.4 & 3.5 & 0.4 & 3.5\\
\textbf{\hspace{1em}2} & \textbf{43.6} & \textbf{21.4} & \textbf{48.9} & \textbf{22.4} & \textbf{50.6} & \textbf{22.7} & \textbf{43.6} & \textbf{21.4} & \textbf{48.9} & \textbf{22.4} & \textbf{48.9} & \textbf{22.4} & \textbf{50.6} & \textbf{22.7}\\
\textbf{\hspace{1em}3} & \textbf{51.7} & \textbf{23.3} & \textbf{49.9} & \textbf{23.7} & \textbf{48.3} & \textbf{23.7} & \textbf{51.7} & \textbf{23.3} & \textbf{49.9} & \textbf{23.7} & \textbf{49.9} & \textbf{23.7} & \textbf{48.3} & \textbf{23.7}\\
\hspace{1em}4 & 3.1 & 20.2 & 0.6 & 20.2 & 0.5 & 20.2 & 3.1 & 20.2 & 0.6 & 20.2 & 0.6 & 20.2 & 0.5 & 20.2\\
\hspace{1em}5 & 1.2 & 8.2 & 0.2 & 8.4 & 0.2 & 8.4 & 1.2 & 8.2 & 0.2 & 8.4 & 0.2 & 8.4 & 0.2 & 8.4\\
\hspace{1em}6 & 0.0 & 2.9 & 0.0 & 2.9 & 0.0 & 2.9 & 0.0 & 2.9 & 0.0 & 2.9 & 0.0 & 2.9 & 0.0 & 2.9\\
\hspace{1em} (None, N) & 0.0 & 79.4 & 0.0 & 81.1 & 0.0 & 81.4 & 0.0 & 79.5 & 0.0 & 81.1 & 0.0 & 81.1 & 0.0 & 81.4\\
\addlinespace[0.3em]
\hline
\multicolumn{15}{c}{\textbf{Scenario 9}}\\
\hspace{1em}1 & 0.0 & 3.3 & 0.0 & 3.3 & 0.0 & 3.3 & 0.0 & 3.3 & 0.0 & 3.3 & 0.0 & 3.3 & 0.0 & 3.3\\
\hspace{1em}2 & 0.0 & 12.9 & 0.2 & 13.5 & 0.3 & 21.9 & 0.0 & 12.9 & 0.2 & 13.5 & 0.2 & 13.5 & 0.3 & 21.9\\
\textbf{\hspace{1em}3} & \textbf{17.9} & \textbf{14.3} & \textbf{83.2} & \textbf{23.2} & \textbf{80.1} & \textbf{23.2} & \textbf{17.9} & \textbf{14.3} & \textbf{79.3} & \textbf{22.7} & \textbf{83.2} & \textbf{23.2} & \textbf{80.1} & \textbf{23.2}\\
\hspace{1em}4 & 18.8 & 23.9 & 2.3 & 23.9 & 2.5 & 23.9 & 18.9 & 24.0 & 3.1 & 24.0 & 2.4 & 24.0 & 2.6 & 24.0\\
\hspace{1em}5 & 25.0 & 22.9 & 4.8 & 23.9 & 5.7 & 24.0 & 25.1 & 22.9 & 5.6 & 23.9 & 4.9 & 23.9 & 5.8 & 24.0\\
\hspace{1em}6 & 38.3 & 22.7 & 9.5 & 23.0 & 11.4 & 23.8 & 38.1 & 22.7 & 11.8 & 23.0 & 9.3 & 23.0 & 11.2 & 23.8\\
\hspace{1em} (None, N) & 0.0 & 100.0 & 0.0 & 110.9 & 0.0 & 120.1 & 0.0 & 100.1 & 0.0 & 110.3 & 0.0 & 110.9 & 0.0 & 120.1\\
\addlinespace[0.3em]
\hline
\multicolumn{15}{c}{\textbf{Scenario 10}}\\
\hspace{1em}1 & 0.0 & 9.2 & 0.0 & 9.2 & 0.0 & 9.2 & 0.0 & 9.2 & 0.0 & 9.2 & 0.0 & 9.2 & 0.0 & 9.2\\
\hspace{1em}2 & 0.0 & 6.2 & 0.0 & 6.2 & 0.0 & 6.2 & 0.0 & 6.2 & 0.0 & 6.2 & 0.0 & 6.2 & 0.0 & 6.2\\
\hspace{1em}3 & 0.0 & 3.5 & 0.0 & 3.5 & 0.0 & 3.5 & 0.0 & 3.5 & 0.0 & 3.5 & 0.0 & 3.5 & 0.0 & 3.5\\
\hspace{1em}4 & 0.2 & 1.6 & 0.2 & 1.6 & 0.2 & 1.6 & 0.2 & 1.6 & 0.2 & 1.6 & 0.2 & 1.6 & 0.2 & 1.6\\
\hspace{1em}5 & 0.0 & 0.9 & 0.0 & 0.9 & 0.0 & 0.9 & 0.0 & 0.9 & 0.0 & 0.9 & 0.0 & 0.9 & 0.0 & 0.9\\
\hspace{1em}6 & 0.0 & 0.7 & 0.0 & 0.7 & 0.0 & 0.7 & 0.0 & 0.7 & 0.0 & 0.7 & 0.0 & 0.7 & 0.0 & 0.7\\
\hspace{1em} (None, N) & \textbf{99.8} & \textbf{22.0} & \textbf{99.8} & \textbf{22.0} & \textbf{99.8} & \textbf{22.0} & \textbf{99.8} & \textbf{22.0} & \textbf{99.8} & \textbf{22.0} & \textbf{99.8} & \textbf{22.0} & \textbf{99.8} & \textbf{22.0}\\
\bottomrule
\end{tabular}
\endgroup{}
\end{table}

\end{document}